\documentclass[usenatbib]{mn2e}
\usepackage[pdftex]{graphicx}
\newcommand{\be}{\begin{equation}}
\newcommand{\ee}{\end{equation}}
\newcommand{\bea}{\begin{eqnarray}}
\newcommand{\eea}{\end{eqnarray}}
\newcommand{\bc}{\begin{center}}
\newcommand{\ec}{\end{center}}

\def\gsim{ \lower .75ex \hbox{$\sim$} \llap{\raise .27ex \hbox{$>$}} }
\def\lsim{ \lower .75ex \hbox{$\sim$} \llap{\raise .27ex \hbox{$<$}} }

\usepackage{afterpage}

\usepackage{amssymb}

\usepackage{times}

\usepackage{paralist}

\usepackage[bookmarks,bookmarksnumbered,colorlinks=true,citecolor=blue,linkcolor=black]{hyperref}
\usepackage{color}

\usepackage{pbox}

\usepackage{booktabs}

\usepackage[flushleft]{threeparttable}

\definecolor{darkgreen}{rgb}{0.0, 0.5, 0.0}

\newcommand{\Msun}{{\rm M_{\odot}}}

\renewcommand{\thefootnote}{\fnsymbol{footnote}}

\title[Stellar mass assembly in Illustris]{The stellar mass assembly of galaxies in the Illustris simulation: growth by mergers and the spatial distribution of accreted stars}

\author[V. Rodriguez-Gomez et al.]
{
	\parbox{18cm}{
		Vicente Rodriguez-Gomez,$^{1 \star}$
		Annalisa Pillepich,$^{1}$
		Laura V. Sales,$^{1,2}$
		Shy Genel,$^{3 \dagger}$ \\
		Mark Vogelsberger,$^{4}$
		Qirong Zhu,$^{5,6}$
		Sarah Wellons,$^{1}$
		Dylan Nelson,$^{1,7}$
		Paul Torrey,$^{4,8}$ \\
		Volker Springel,$^{9,10}$
		Chung-Pei Ma,$^{11}$
		and Lars Hernquist$^{1}$
	}
	\vspace{0.3cm} \\ 
	$^{1}$ Harvard-Smithsonian Center for Astrophysics, 60 Garden Street, Cambridge, MA 02138, USA \\
	$^{2}$ Department of Physics \& Astronomy, University of California, Riverside, 900 University Avenue, Riverside, CA 92521, USA \\
	$^{3}$ Department of Astronomy, Columbia University, 550 West 120th Street, New York, NY 10027, USA \\
	$^{4}$ Department of Physics, Kavli Institute for Astrophysics and Space Research, Massachusetts Institute of Technology, Cambridge, MA 02139, USA \\
  	$^{5}$ Department of Astronomy \& Astrophysics, The Pennsylvania State University, 525 Davey Lab, University Park, PA 16802, USA \\
  	$^{6}$ Institute for Gravitation and the Cosmos, The Pennsylvania State University, University Park, PA 16802, USA \\
  	$^{7}$ Max-Planck-Institut f\"{u}r Astrophysik, Karl-Schwarzschild-Stra\ss{}e 1, 85741 Garching bei M\"{u}nchen, Germany \\
  	$^{8}$ TAPIR, Mailcode 350-17, California Institute of Technology, Pasadena, CA 91125, USA \\
	$^{9}$ Heidelberg Institute for Theoretical Studies, Schloss-Wolfsbrunnenweg 35, 69118 Heidelberg, Germany \\
	$^{10}$ Zentrum f\"{u}r Astronomie der Universit\"{a}t Heidelberg, ARI, M\"onchhofstr. 12-14, 69120 Heidelberg, Germany \\
	$^{11}$ Department of Astronomy, University of California, Berkeley, CA 94720, USA
}

\begin{document}

%\tableofcontents

\maketitle
\begin{abstract}
We use the Illustris simulation to study the relative contributions of in situ star formation and stellar accretion to the build-up of galaxies over an unprecedentedly wide range of masses ($M_{\ast} = 10^9-10^{12} \, {\rm M_{\odot}}$), galaxy types, environments, and assembly histories. We find that the `two-phase' picture of galaxy formation predicted by some models is a good approximation only for the most massive galaxies in our simulation -- namely, the stellar mass growth of galaxies below a few times $10^{11} \, {\rm M_{\odot}}$ is dominated by in situ star formation at all redshifts. The fraction of the total stellar mass of galaxies at $z=0$ contributed by accreted stars shows a strong dependence on galaxy stellar mass, ranging from about 10 per cent for Milky Way-sized galaxies to over 80 per cent for $M_{\ast} \approx 10^{12} \, {\rm M_{\odot}}$ objects, yet with a large galaxy-to-galaxy variation. At a fixed stellar mass, elliptical galaxies and those formed at the centres of younger haloes exhibit larger fractions of ex situ stars than disc-like galaxies and those formed in older haloes. On average, $\sim$50 per cent of the ex situ stellar mass comes from major mergers (stellar mass ratio $\mu > 1/4$), $\sim$20 per cent from minor mergers ($1/10 < \mu < 1/4$), $\sim$20 per cent from very minor mergers ($\mu < 1/10$), and $\sim$10 per cent from stars that were stripped from surviving galaxies (e.g. flybys or ongoing mergers). These components are spatially segregated, with in situ stars dominating the innermost regions of galaxies, and ex situ stars being deposited at larger galactocentric distances in order of decreasing merger mass ratio.
\end{abstract}


\section{Introduction}\label{sec:intro}
\renewcommand{\thefootnote}{\fnsymbol{footnote}}
\footnotetext[1]{E-mail: vrodriguez-gomez@cfa.harvard.edu}
\footnotetext[2]{Hubble fellow.}

% Numbered footnotes from now on:
\renewcommand{\thefootnote}{\arabic{footnote}}

In the $\Lambda$ cold dark matter ($\Lambda$CDM) cosmological model, structure forms hierarchically: galaxies grow by accreting smaller systems composed of both dark matter (DM) and baryons. However, the importance of mergers (as opposed to secular processes) in driving the growth of the stellar components of galaxies, as well as in determining their resulting kinematic and chemical properties, are still the subject of significant discussion in the literature, both theoretically \citep[e.g.][]{Hopkins2010, DeLucia2011, Sales2012, Zavala2012, Fiacconi2014} and observationally \citep[e.g.][]{Bundy2007, Oesch2010, Lopez-Sanjuan2012}. In particular, the scientific community has not yet come to an agreement in regards to the amount of stars that were formed in situ (i.e. formed from accreted gas within the galaxy where they are currently found), relative to stars that were formed ex situ (i.e. formed in other galaxies and subsequently accreted) \citep[e.g.][]{Oser2010, Lackner2012a}. Furthermore, theoretical predictions for the spatial distribution of accreted stars remains an ongoing effort \citep[e.g.][]{Deason2013, Pillepich2015, Hirschmann2015} as a powerful means to explain many observed properties of stellar haloes \citep[e.g.][]{Deason2011, Pastorello2014, Greene2015}.

One of the first systematic studies of the stellar mass assembly of galaxies that used hydrodynamic simulations as the main tool was carried out by \cite{Oser2010}, who performed zoom-in simulations of 39 haloes with virial masses between $7.0 \times 10^{11} h^{-1}\Msun$ and $2.7 \times 10^{13} h^{-1}\Msun$, and used them to investigate galaxy formation by distinguishing stars that were formed in situ from those that were accreted. They found that their simulated galaxies exhibit `two phases' of stellar mass growth: an early stage of in situ star formation until $z \sim 2$, which resembles so-called `monolithic collapse' models \citep[e.g.][]{Eggen1962}, followed by a dry merger-dominated stage with markedly smaller star formation rates.

In general, \cite{Oser2010} found very high ex situ stellar mass fractions for all of the galaxies they studied, with values ranging from 60 to 80 per cent with increasing stellar mass. Similarly, \cite{Lackner2012a} studied the in situ and ex situ stellar populations of 611 galaxies formed in two hydrodynamic zoom-in simulations, one of them centred on a massive cluster, and the other one centred on a void. For galaxies with $M_{\ast} \approx 10^{11} \Msun$, they found ex situ stellar mass fractions which are lower by a factor of $\sim$3 than those found by \cite{Oser2010}, which implies that accreted stars are no longer the dominant stellar component for galaxies of this mass.

Similar studies have been carried out with semi-analytic models (SAMs) of galaxy formation \citep[e.g.][]{Guo2008, Jimenez2011, Lee2013}. In particular, \cite{Lee2013} determined that the ex situ stellar mass fraction is a strong function of stellar mass, finding values of approximately 20, 40, and 70 per cent for galaxies in the mass ranges $\log\left(M_{\ast}/\Msun\right) =$ 10.5--11.0, 11.0--11.5, and 11.5--12.0, respectively. \cite{Lee2013} also quantified the amount of merger-induced `bursty' star formation, which they found to be negligible compared to quiescent, `disc-mode' star formation, in agreement with \cite{Hopkins2010a}. This is explained by the fact that most gas-rich mergers happen at high redshifts when galaxies are smaller, while recent mergers tend to be gas-poor and happen more rarely.

Investigating the in situ and ex situ stellar components of simulated galaxies can provide important insights into galaxy formation and the effects from galaxy interactions. For instance, the transformation of disc to elliptical galaxies is believed to take place by the addition of ex situ stars from dry mergers, which are deposited at large radii. The size evolution of massive, compact galaxies observed at $z \sim 2$ \citep[e.g.][]{Daddi2005, Trujillo2006, VanDokkum2008, Damjanov2009} provides an example of such transformation, since many of them are believed to eventually become the `cores' of large, massive ellipticals in the local Universe \citep[e.g.][]{Naab2009, Hopkins2010b, Feldmann2010, VanDokkum2010, Cimatti2012, Oser2012, Wellons2016}.

Other studies have focused on the stellar haloes around simulated galaxies, either using hydrodynamic simulations of large samples of galaxies \citep{Font2011, McCarthy2012, Pillepich2014} or highly resolved individual galaxies \citep{Abadi2006, Zolotov2009, Zolotov2010, Tissera2012, Tissera2013, Tissera2014, Dubois2013, Pillepich2015, Hirschmann2015}, or using $N$-body simulations along with particle tagging techniques \citep{Cooper2010, Cooper2013, Cooper2015, Rashkov2012}. Such theoretical predictions for the spatial distribution of in situ and ex situ stars can be used to explain several observational features of stellar haloes at large galactocentric distances, such as metallicity and stellar age gradients, abundance ratios, kinematics and velocity anisotropy profiles, or degree of substructure. With the advent of next-generation deep and wide-field surveys, it will become necessary to have reliable theoretical predictions that will aid in guiding and explaining observations.

In the present work we provide a comprehensive view of the stellar mass assembly of galaxies using the Illustris simulation \citep{Vogelsberger2014, Vogelsberger2014a, Genel2014a, Nelson2015}, a hydrodynamic cosmological simulation carried out on a periodic box of $\sim$106.5 Mpc per side, which features a realistic physical model \citep{Vogelsberger2013, Torrey2014}. Because of the wide range of stellar masses, environments, and spatial scales covered, the Illustris simulation presents a unique opportunity to study the assembly of the different stellar components of galaxies in a self-consistent, cosmological setting.

This paper is organized as follows. In Section \ref{sec:methodology} we describe the Illustris simulation, the merger trees, and the stellar particle classification scheme. In Section \ref{sec:stellar_mass_accretion} we define and quantify the \textit{specific stellar mass accretion rate}, for which we provide a fitting function that is accurate over a wide range of stellar masses, merger mass ratios, and redshifts. The \textit{ex situ stellar mass fraction} is introduced in Section \ref{sec:ex_situ_stellar_mass_fraction}, including a discussion of its general trends at $z=0$ (Section \ref{subsec:general_trends}), its correlation with galaxy properties such as stellar age, morphology and assembly history (Section \ref{subsec:f_acc_by_type}), and its redshift evolution (Section \ref{subsec:redshift_evolution}). In Section \ref{sec:spatial_distribution} we examine the spatial distribution of ex situ stars and present a systematic study of the \textit{normalized transition radius}. Finally, we discuss our results and present our conclusions in Section \ref{sec:discussion_and_conclusions}.

\section{Methodology}\label{sec:methodology}

\subsection{The Illustris simulation}\label{subsec:illustris}

Throughout this paper we use data from the Illustris Project \citep{Vogelsberger2014, Vogelsberger2014a, Genel2014a}, a suite of hydrodynamic cosmological simulations of a periodic box with 106.5 Mpc on a side, carried out with the moving-mesh code \textsc{arepo} \citep{Springel2010}. Illustris-1 (known hereafter as \textit{the} Illustris simulation) follows the joint evolution of $1820^3$ DM particles along with approximately $1820^3$ gas cells or stellar particles. Each DM particle has a mass of $6.26 \times 10^6 \Msun$, while the average mass of the baryonic elements is $1.26 \times 10^6 \Msun$. The gravitational softening for DM particles is 1.4 kpc in comoving coordinates, while for stellar particles this scale is held constant at 0.7 kpc in physical coordinates at $z < 1$. The Illustris-2 and Illustris-3 simulations are lower-resolution versions of Illustris-1, carried out with $2 \times 910^3$ and $2 \times 455^3$ resolution elements, respectively.

The Illustris simulation features a galaxy formation model which includes star formation and evolution, primordial and metal-line cooling, chemical enrichment, gas recycling, and feedback from supernovae and super-massive black holes. This galaxy formation model is described in detail in \cite{Vogelsberger2013} and \cite{Torrey2014}, where it is also shown to reproduce several key observables across different redshifts. The model has also been found to be in good agreement with a number of observables for which it was not tuned, including the column density distribution of neutral hydrogen \citep[e.g.][]{Bird2014}, the properties of dwarf and satellite galaxies around massive hosts \citep{Sales2014a, Mistani2016}, and the observed galaxy merger rate \citep{Rodriguez-Gomez2015}. 

The main output from the Illustris simulation consists of 136 snapshots between $z=46$ and $z=0$. The first 75 snapshots (at $z > 3$) are spaced logarithmically in the cosmic scale factor $a$, while the other 61 (at $z < 3$) have a finer time spacing of $\Delta t \approx 0.15$ Gyr. For each of these snapshots, DM haloes are identified by applying the friends-of-friends (FoF) algorithm \citep{Davis1985} with a linking length equal to 0.2 times the mean interparticle separation (each baryonic element being assigned to the same FoF group as the closest DM particle). Within each FoF group, substructure is identified using an updated version of the \textsc{subfind} algorithm \citep{Springel2001, Dolag2009a} which can be applied to hydrodynamic simulations, so that every FoF group is associated to one \textit{central} \textsc{subfind} halo and possibly a family of satellite objects. Throughout this paper, FoF groups are referred to as \textit{haloes} and \textsc{subfind} haloes as \textit{subhaloes}, unless noted otherwise. Moreover, we define a galaxy as the baryonic component of a \textsc{subfind} halo (central or satellite) with $M_{\ast}>0$. At any given time, stellar particles and gas elements belong to a given galaxy if they are gravitationally bound to such galaxy according to the \textsc{subfind} algorithm, regardless of their distance to the galactic centre. In all computations we will characterize galaxies according to their total stellar mass, rather than by using the stellar mass measured within some fiducial aperture such as twice the stellar half-mass radius.

The Illustris simulation was carried out with a $\Lambda$CDM cosmological model with parameters $\Omega_{\rm m} = 0.2726$, $\Omega_{\Lambda} = 0.7274$, $\Omega_{\rm b} = 0.0456$, $\sigma_8 = 0.809$, $n_{\rm s} = 0.963$, and $h = 0.704$, in agreement with the 9-yr \textit{Wilkinson Microwave Anisotropy Probe (WMAP)} results \citep{Hinshaw2013}.

\subsection{Merger trees}\label{subsec:merger_trees}

Merger trees of the subhaloes have been constructed with the \textsc{sublink} algorithm \citep{Rodriguez-Gomez2015}, which proceeds in two main stages: (1) finding subhalo descendants and (2) rearranging this information in order to make it `usable' for galaxy formation studies (i.e. constructing the merger trees). For the first stage, each subhalo from a given snapshot is assigned a \textit{unique descendant} from the next snapshot by comparing particle IDs in a weighted fashion, assigning a higher priority to particles that are more tightly bound. In some special cases, a subhalo is allowed to `skip' a snapshot when finding a descendant, which accounts for situations in which a small subhalo is temporarily `lost' due to insufficient density contrast while it is traversing a larger structure. Once all descendants have been determined, the \textit{first progenitor} (also known as the \textit{main progenitor}) is defined as the one with the `most massive history' behind it \citep{DeLucia2007}. In this paper we make exclusive use of the \textit{baryonic} merger trees, constructed by tracking only the stellar particles and star-forming gas cells of subhaloes. Differences with respect to the \textit{DM-only} merger trees and further details about the algorithm can be found in \cite{Rodriguez-Gomez2015}, while a complete description of the data format is presented in \cite{Nelson2015}.

\subsection{Stellar particle classification}\label{subsec:stellar_particle_classification}

Although the merger trees by themselves are a very effective tool for studying galaxy formation and evolution, further insight can be obtained by individually classifying stellar particles in the simulation according to their formation and accretion histories. To this end, we have created a stellar assembly catalogue which contains the following information for every stellar particle from every snapshot:

\begin{itemize}
  \item \textit{In situ / ex situ:} A stellar particle is considered to have been formed \textit{in situ} if the galaxy in which it formed lies along the `main progenitor branch' (in the merger trees) of the galaxy in which it is currently found. Otherwise, the stellar particle is tagged as \textit{ex situ}.\\
  
  \item \textit{After infall / before infall:} An ex situ stellar particle is classified as \textit{after infall} if the galaxy in which it formed was already part of the halo (FoF group) where it is currently found. Otherwise, the ex situ stellar particle is labelled \textit{before infall}. \footnote{In the case of multiple halo passages, this definition refers to the \textit{last} time the galaxy entered the halo under consideration.} \\
  
  \item \textit{Stripped from surviving galaxies / accreted through mergers:} An ex situ stellar particle belongs to the \textit{stripped from surviving galaxies} category if it has been stripped from the galaxy in which it formed and this galaxy has not merged with the galaxy where the particle is currently found. Otherwise, the ex situ stellar particle is classified as \textit{accreted through mergers}.\\
  
  \item \textit{Merger mass ratio:} For each stellar particle accreted through completed mergers, this is the stellar mass ratio of the merger in which the particle was accreted, measured at the moment when the galaxy in which the particle formed reaches its maximum stellar mass.
\end{itemize}

A stellar particle in the \textit{stripped from surviving galaxy} category must have been `captured' during a flyby event or during a close passage of an ongoing merger. We will not distinguish between these different scenarios, and we note that a stellar particle in this category at one time can eventually be reclassified as \textit{accreted through mergers} at a later time if the galaxy in which the stellar particle formed eventually merges with the galaxy where it is currently found.

We emphasize that the definitions of in situ and ex situ stars can vary substantially in the literature. As detailed in Section \ref{subsec:illustris}, stellar and gas elements are considered to belong to a given galaxy according to the \textsc{subfind} algorithm and its unbinding procedure. However, our classification scheme is based exclusively on the origin of the stellar particles, without additional considerations on the gas out of which the stars form, differently from other works. In fact, we note that the gas element progenitors of in situ stars of a given galaxy can have become bound to it either through smooth gas accretion onto the parent halo, via mergers, or through tidal and ram pressure stripping of gas from other nearby galaxies. In this work we do not distinguish between these gas progenitor channels. We also note that, in the case of central galaxies, our ex situ category does \textit{not} include stars which formed in satellite galaxies but which have not been stripped. Therefore, our ex situ stellar mass does \textit{not} include satellite stellar mass.

Finally, we point out that the fraction of stars in Illustris which were not bound to any \textsc{subfind} halo at formation (e.g. from gas cells turning into stars along gas filaments or clumps outside any identified galaxy) is very small, namely 0.1 per cent at $z=0$. On the other hand, ex situ stars in Illustris rarely accrete \textit{smoothly} into their final host galaxy. We estimate that only 0.2--0.3 per cent of the ex situ stars have been accreted onto galaxies without being bound to any \textsc{subfind} halo, possibly having been kicked far away from their birth sites as the result of a violent merger.

\subsection{The galaxy sample}

We consider all galaxies from the Illustris simulation with stellar masses $M_{\ast} > 10^{9} \, \Msun$, without imposing any further restrictions. This choice produces a sample of 29203 objects at $z=0$, including both centrals and satellites (no distinction is made between the two classes except when explicitly stated). For some calculations in the next sections, results are shown for three stellar mass bins centred at $10^{10}$, $10^{11}$, and $10^{12} \, \Msun$ (in log-space), each with a full width equivalent to a factor of $10^{1/3} \approx 2.15$, and therefore containing 3829, 804, and 37 galaxies, respectively.

\section{The specific stellar mass accretion rate}\label{sec:stellar_mass_accretion}

Before exploring the in situ and ex situ (accreted) stellar components of galaxies in Section \ref{sec:ex_situ_stellar_mass_fraction}, we dedicate the current section to quantifying the \textit{instantaneous} rate of mass growth due to stellar accretion. In particular, we introduce the \textit{specific stellar mass accretion rate}, which measures the average amount of stellar mass that a galaxy accretes per unit time through mergers with other galaxies. The specific stellar mass accretion rate is an interesting quantity on its own, as it can be directly compared to the specific star formation rate \citep[e.g.][]{Moster2012a, Behroozi2013b}, the specific DM accretion rate \citep[e.g.][figure 13]{Genel2014a}, or the cosmological gas accretion rate \citep[e.g.][]{Keres2005, Dekel2009, Nelson2013}, in order to evaluate the importance of dissipative processes with respect to galaxy formation. In addition, the specific stellar mass accretion rate acts as a link between the galaxy-galaxy merger rate studied in \cite{Rodriguez-Gomez2015} and the ex situ stellar mass fraction that will be presented in Section \ref{sec:ex_situ_stellar_mass_fraction}.

In the current section we define the specific stellar mass accretion rate, discuss some of its basic properties, and provide a fitting function that is accurate over a wide range of stellar masses, merger mass ratios, and redshifts. We also evaluate the importance of mergers relative to star formation when building up the stellar mass of a galaxy. Throughout this section, our galaxy sample is extended down to a minimum stellar mass of $10^{8} \, \Msun$.

\subsection{Definitions}\label{subsec:macc_definitions}

\begin{figure*}
	\centerline{
		\vbox{
			\includegraphics[width=17.5cm]{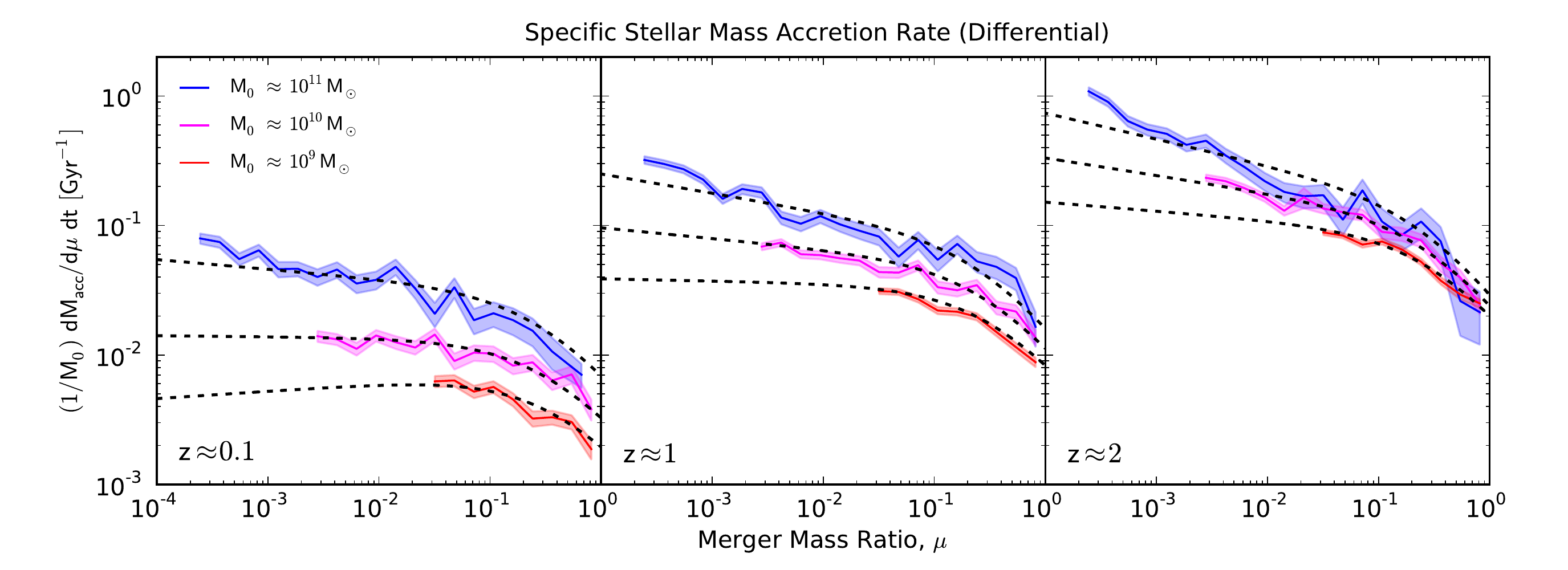}
			\includegraphics[width=17.5cm]{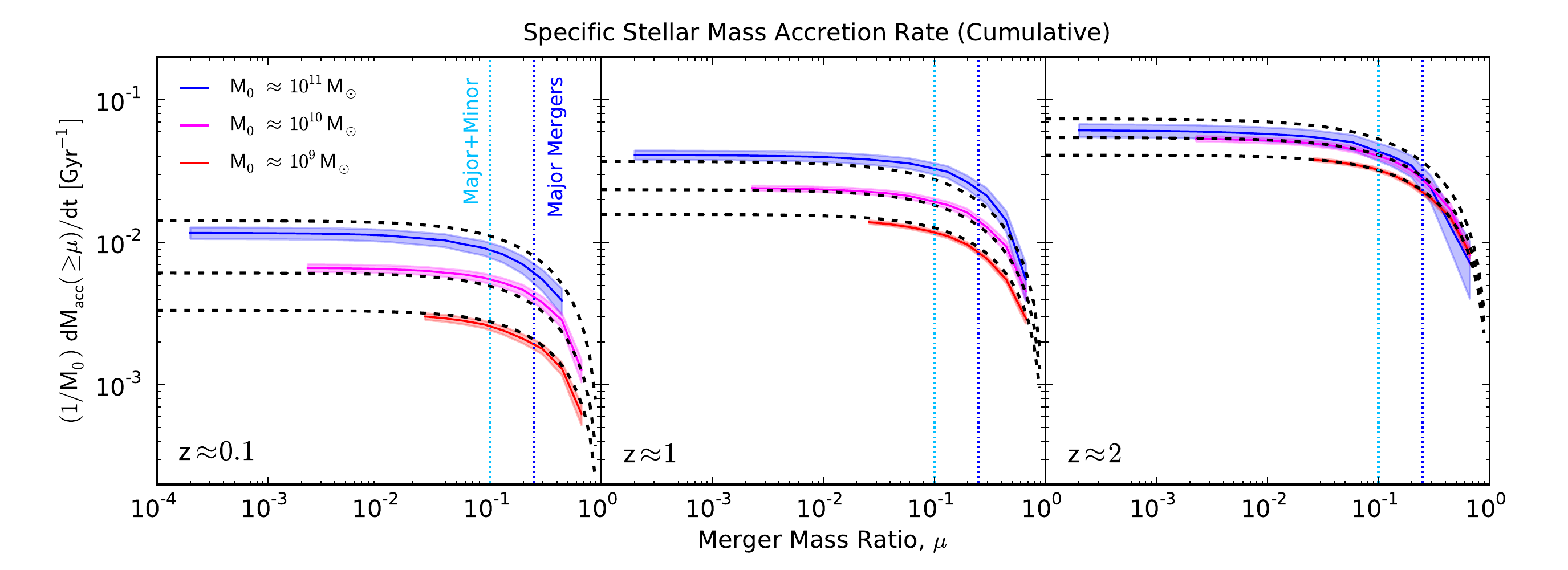}
		}
	}
\caption{The specific stellar mass accretion rate as a function of merger mass ratio $\mu$, shown for different descendant masses $M_0$. Each mass bin is a factor of $\sim$2 wide. The left, centre, and right panels correspond to redshift bins centred around $z = 0.1$, 1, and 2, respectively. In all panels, the shaded regions correspond to the Poisson noise in the number of mergers, while the black dashed line represents the fitting function from Table \ref{tab:fitting_formula}. Top: the \textit{differential} specific stellar mass accretion rate, as given by equation (\ref{eq:stellar_mass_accretion_rate}). Bottom: the \textit{cumulative} specific stellar mass accretion rate, i.e. integrated with respect to merger mass ratio $\mu$ so that each $y$-value represents the growth rate due to all mergers with mass ratios greater than the value given on the $x$-axis (note that in this case the dashed black line is actually the \textit{integral} of the fitting function with respect to $\mu$, and therefore it is not a direct fit to the data). For reference, the growth rates due to major ($\mu > 1/4$) and major + minor ($\mu > 1/10$) mergers are given by the intersections with the vertical blue and cyan dotted lines, respectively. The fact that each curve in the lower panels asymptotes very quickly towards low merger mass ratios demonstrates the decreased importance of mergers with $\mu \lesssim 1/100$ with respect to galaxy growth, despite the fact that we can actually resolve mergers with even smaller mass ratios.}
\label{fig:macc_vs_ratio}
\end{figure*}

Following the merger analysis presented in \cite{Rodriguez-Gomez2015}, we define galaxy mergers in the following way. If a galaxy in the merger trees has $N_{\rm p}$ direct progenitors, we count $N_{\rm p} - 1$ mergers, occurring between the first progenitor and each of the other ones. For each merger, we denote the stellar masses of the descendant, primary progenitor and secondary progenitor by $M_0$, $M_1$ and $M_2$, respectively. Both progenitor masses $M_1$ and $M_2$ are measured at the time when $M_2$ reaches its maximum value, which we refer to as $t_{\rm max}$. The corresponding merger mass ratio is given by $\mu = M_2 / M_1$. In the case of multiple mergers ($N_{\rm p} > 2$), the primary progenitor is always the same object, but its stellar mass $M_1$ is generally measured at a different time for each merger, i.e. at the time when $M_2$ reaches its maximum value.

With the definitions above, we can introduce the \textit{specific stellar mass accretion rate}, which measures the \textit{average} amount of stellar mass accreted by a single `descendant' galaxy during a time interval ${\rm d}t$, normalized by the stellar mass of the descendant galaxy $M_0$, and considering only mergers with stellar mass ratios inside the range $(\mu, \mu+{\rm d}\mu)$. The specific stellar mass accretion rate, which we denote by $\dot{m}_{\rm acc, \ast}$, is a function of descendant stellar mass $M_0$, merger mass ratio $\mu$, and redshift $z$:

\begin{equation}
  \dot{m}_{\rm acc, \ast}(M_0, \mu, z) = \frac{1}{M_0}\frac{{\rm d}{M}_{\rm acc}}{{\rm d}\mu \, {\rm d}t},
	\label{eq:stellar_mass_accretion_rate}
\end{equation}
where
\[
M_{\rm acc} = \sum_{k=1}^{N_{\rm p}-1} M_{2,k}
\]
is the total stellar mass contributed by the secondary progenitors\footnote{More precisely, we add $\min(M_1, M_2)$ for each merger, consistent with the condition that the mass ratio $\mu$ should always be smaller than one \citep[see][for a discussion]{Rodriguez-Gomez2015}.} of the mergers that took place in the appropriate $(M_0, \mu, z)$ interval. Each stellar mass contribution $M_{2,k}$ (corresponding to a merger of stellar mass ratio $\mu_{k} = M_{2,k} / M_1$) is also measured at $t_{\rm max}$. The specific stellar mass accretion rate is very closely related to the galaxy-galaxy merger rate: it is also a global, mean quantity, giving the average accretion rate \textit{per galaxy}, and is also given in units of Gyr$^{-1}$. We note that the specific stellar mass accretion rate given by equation (\ref{eq:stellar_mass_accretion_rate}) is mathematically similar to the dimensionless growth rate studied by \cite{Guo2008}, with the important difference that in their work the time interval ${\rm d}t$ is always normalized by the age of the Universe $t(z)$. Additionally, our specific stellar mass accretion rate is given as a function of stellar mass ratio $\mu$ and can therefore be integrated over any $\mu$-interval in order to evaluate the importance of major mergers, minor mergers, or all mergers (i.e. by integrating from $\mu=0$ to $\mu=1$).

In practice, we approximate equation (\ref{eq:stellar_mass_accretion_rate}) by going through the following five steps: (1) defining small three-dimensional bins in $(M_0, \mu, z)$-space (yet large enough to contain at least 5 merger events each), (2) adding the stellar mass contributions (from the secondary progenitors) of all mergers that take place in each bin, (3) dividing by the number of descendant galaxies in each bin, (4) dividing by the descendant stellar mass $M_0$, and finally (5) dividing by the appropriate time and merger mass ratio intervals (corresponding to the edges of the selected bins).

\subsection{Connection with the galaxy merger rate}\label{subsec:connection_with_merger_rate}

The merger rate of \textit{DM haloes} has been systematically studied in previous work using $N$-body simulations of structure formation in the Universe, with different predictions showing agreement within a factor of $\sim$1.5 \citep{Fakhouri2008, Fakhouri2010, Genel2009, Genel2010}. Furthermore, the merger rate of \textit{galaxies} can be estimated using semi-empirical methods, which combine results from $N$-body simulations with observational constraints \citep[e.g.][]{Stewart2009, Hopkins2010}, and it has been recently determined to great accuracy using cosmological hydrodynamic simulations in \cite{Rodriguez-Gomez2015}. In the latter study, the galaxy-galaxy merger rate was found to have a relatively simple mathematical form as a function of stellar mass, merger mass ratio, and redshift.

In principle, it seems possible to calculate the specific stellar mass accretion rate based on the galaxy-galaxy merger rate alone. In practice, however, this is complicated by the fact that the mass ratio of a merger must be measured at an earlier time than when the merger actually happens, such as $t_{\rm max}$, i.e. the time when the secondary progenitor reaches its maximum stellar mass \citep[see][for a discussion]{Rodriguez-Gomez2015}. Between $t_{\rm max}$ and the moment when the merger finally takes place, one or both progenitors can undergo significant star formation or have interactions with a third object. This implies that the stellar mass conservation relation $M_0 = M_1 + M_2$ is not satisfied in general. Therefore, it becomes necessary to measure the specific stellar mass accretion rate directly from the simulation, rather than attempting to derive it from the merger rate.

If we assume that the stellar mass conservation relation $M_0 = M_1 + M_2$ is satisfied, then the specific stellar mass accretion rate (equation \ref{eq:stellar_mass_accretion_rate}) and the galaxy-galaxy merger rate would be related by 
	\begin{equation}
		\frac{1}{M_0}\frac{{\rm d}M_{\rm acc}}{{\rm d}\mu \, {\rm d}t} = \frac{\mu}{1+\mu}\frac{{\rm d}N_{\rm mergers}}{{\rm d}\mu \, {\rm d}t}.
		\label{eq:conversion}
	\end{equation}
Essentially, the only difference between the two quantities would be a factor of $\mu / (1+\mu)$. In practice, however, and given that stellar mass is generally not conserved during a merger, we find that a `conversion' factor of $\mu / (1 + 3 \mu)$ results in a better description of the data (see Section \ref{subsec:macc_fitting_formula}). This empirical correction down-weighs the major-merger end in the right-hand side of equation (\ref{eq:conversion}) by a factor of $\sim$2, suggesting that in this regime the primary progenitor can grow by a similar amount between $t_{\rm max}$ and the time of the merger, as a consequence of in situ star formation and mergers with additional objects.

\subsection{Results}\label{subsec:macc_results}

\begin{figure*}
	\centerline{
		\vbox{
			\includegraphics[width=17.5cm]{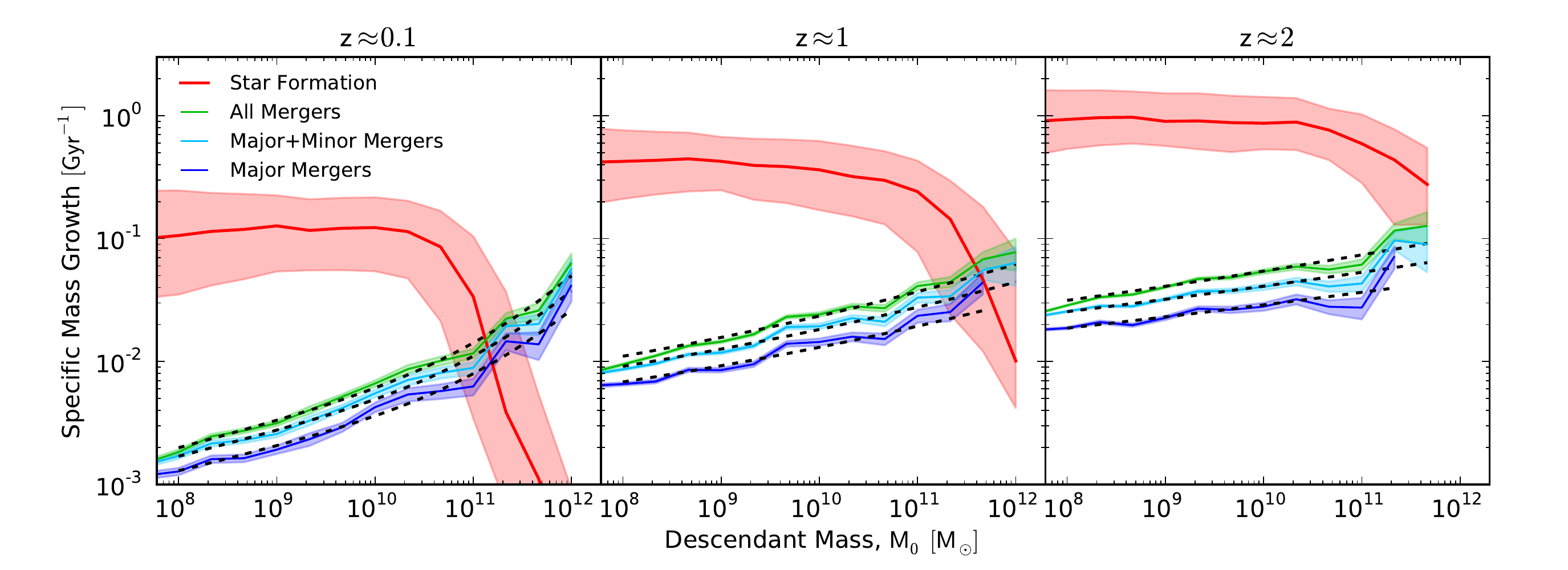}
		}
	}
\caption{The \textit{cumulative} (with respect to merger mass ratio $\mu$) specific stellar mass accretion rate as a function of descendant mass $M_0$, shown for major (blue, $\mu > 1/4$), major + minor (cyan, $\mu > 1/10$), and all mergers (green). The left, centre, and right panels correspond to redshift bins centred around $z = 0.1$, 1, and 2, respectively. In all panels, the shaded regions correspond to the Poisson noise in the number of mergers, while the black dashed line represents the fitting function from Table \ref{tab:fitting_formula}, integrated over the appropriate range in merger mass ratio $\mu$. The red line shows the median specific star formation rate for each mass bin, with the shaded region around it indicating the range between the 16th and 84th percentiles, or $1\sigma$. This figure shows that the specific stellar mass accretion rate has a relatively strong dependence on stellar mass, and also that the stellar mass growth of most galaxies is dominated by in situ star formation rather than stellar accretion from mergers, except for sufficiently massive galaxies at $z \lesssim 1$.}
\label{fig:macc_vs_mass}
\end{figure*}

	\begin{table*}
		\centering

		\begin{tabular}{c | c | c}
			\hline
			Definition & $\dot{m}_{\rm acc, \ast}(M_0, \mu, z) = \frac{1}{M_0}\frac{{\rm d}{M}_{\rm acc}}{{\rm d}\mu \, {\rm d}t}$ \\
			\addlinespace[1ex]
			Units & Gyr$^{-1}$ \\
			\hline
			\parbox{2cm}{Fitting function} &
			\parbox{9cm}{\centering $A(z) \, \left(\frac{M_0}{10^{10}\Msun}\right)^{\alpha(z)} \left[1 +  \left(\frac{M_0}{M_0^{\prime}}\right)^{\delta(z)}\right] \mu^{\beta (z) + \gamma \log_{10}\left(\frac{M_0}{10^{10}\Msun}\right)} \left(\frac{\mu}{1 + 3\mu}\right)$, \\
			                               where \\
			                               $A(z) = A_0 (1+z)^{\eta}$, \\
			                               $\alpha(z) = \alpha_0 (1+z)^{\alpha_1}$, \\
			                               $\beta(z) = \beta_0 (1+z)^{\beta_1}$, \\
			                               $\delta(z) = \delta_0 (1+z)^{\delta_1}$, \\
			                               and $M_0^{\prime} = 2 \times 10^{11} \, \Msun$ is fixed.} \\
			\addlinespace[1ex]
			\parbox{2cm}{\centering
				$\log_{10}(A_0/{\rm Gyr^{-1}})$ \\
				$\eta$ \\
				$\alpha_0$ \\
				$\alpha_1$ \\
				$\beta_0$ \\
				$\beta_1$ \\
				$\gamma$ \\
				$\delta_0$ \\
				$\delta_1$} &
			\parbox{4cm}{\flushright
				$-2.0252 \pm 0.0060$ \\
				$1.5996 \pm 0.0146$ \\
				$0.2013 \pm 0.0050$ \\
				$-1.4888 \pm 0.0540$ \\
				$-0.9964 \pm 0.0030$ \\
				$0.1177 \pm 0.0030$ \\
				$-0.0656 \pm 0.0015$ \\
				$0.6949 \pm 0.0311$ \\
				$-1.7581 \pm 0.0675$} \hspace{2cm} \\

			\addlinespace[1ex]
			$\chi^2_{\rm red}$ & 1.21 \\
			\hline
		\end{tabular}

		\caption{Fitting function and best-fitting parameters for the specific stellar mass accretion rate. See Section \ref{subsec:macc_fitting_formula} for details.}
		\label{tab:fitting_formula}
	\end{table*}

Fig. \ref{fig:macc_vs_ratio} shows the specific stellar mass accretion rate as a function of merger mass ratio $\mu$, calculated for different descendant masses $M_0$. The panels from left to right correspond to redshift bins centred around $z = 0.1$, 1, and 2. The upper panels show the specific stellar mass accretion rate in its most general form, as given by equation (\ref{eq:stellar_mass_accretion_rate}), while the lower panels show the \textit{cumulative} mass accretion rate, integrated for mass ratios larger than the value on the $x$-axis. For convenience, the specific stellar mass accretion rates from major ($\mu > 1/4$) and major + minor ($\mu > 1/10$) mergers are indicated with blue and cyan vertical dotted lines, respectively. The shaded regions represent the uncertainty arising from Poisson noise in the number of mergers, while the black dashed line shows predictions from the fitting function presented in Table \ref{tab:fitting_formula} and discussed in Section \ref{subsec:macc_fitting_formula}.

The lower panels of Fig. \ref{fig:macc_vs_ratio} demonstrate the rapidly decreasing importance of mergers with respect to mass ratio. Although we are able to resolve mergers with mass ratios $\mu \sim 10^{-4}$ and below (and find them to be quite numerous), the cumulative effect from mergers with stellar mass ratios below $\mu \sim 1/100$ is essentially negligible with respect to stellar mass growth, as can be seen from the fact that the curves in the lower panels of Fig. \ref{fig:macc_vs_ratio} quickly asymptote toward low values of $\mu$.

In Fig. \ref{fig:macc_vs_mass} we show the \textit{cumulative} (with respect to mass ratio) specific stellar mass accretion rate as a function of descendant stellar mass $M_0$. The blue, cyan and green lines show the contributions from major, major + minor, and all mergers, respectively. The panels from left to right correspond to redshift bins centred around $z = 0.1$, 1, and 2. As before, the shaded regions correspond to the Poisson noise in the number of mergers, while the dashed black lines represent the model from Table \ref{tab:fitting_formula}. For comparison purposes, the median specific star formation rate is shown in red, along with its associated $1 \sigma$ scatter.

It is clear from Figs. \ref{fig:macc_vs_ratio} and \ref{fig:macc_vs_mass} that the contribution to stellar mass growth due to different types of mergers is 50--60 per cent from major mergers, 20--25 per cent from minor mergers, and 20--25 per cent from very minor mergers. Fig. \ref{fig:macc_vs_mass} also shows that the growth of galaxies below a certain stellar mass is dominated by in situ star formation, while more massive galaxies above this threshold grow primarily by ex situ contributions. The value of this transition point decreases from $\sim$5 $\times 10^{11} \Msun$ at $z \approx 1$ to 1--2 $\times 10^{11} \Msun$ at $z \approx 0.1$. Our results are in broad agreement with observational work by \cite{Robotham2014a}, who found a transition between the two modes of stellar mass growth at $M_{\ast} \approx 7 \times 10^{10} \Msun$. In general, these trends are explained by the fact that massive galaxies have mergers more frequently than smaller galaxies, while at the same time their star formation rate is substantially suppressed by AGN feedback \citep[see][for a discussion of star formation and its scatter in Illustris]{Sparre2015}. 

Comparing with \cite{Guo2008}, we find that the specific stellar mass accretion rate has a relatively weak dependence on stellar mass and a strong redshift evolution, whereas Guo \& White found precisely the opposite: a strong dependence on stellar mass and a weak redshift evolution. In other words, we find that the mathematical form of the specific growth rate of galaxies contributed by mergers is more similar to that of DM haloes than previously thought. Given that the stellar mass accretion rate is closely related to the galaxy-galaxy merger rate, these contrasting trends originate from the analogous qualitative differences found in the galaxy-galaxy merger rates from \cite{Guo2008} and \cite{Rodriguez-Gomez2015}, as discussed in the latter work. In spite of these qualitatively different trends with mass and redshift, it is noteworthy that \cite{Guo2008} found that the instantaneous growth rates due to mergers and star formation become comparable at $z=0$ for galaxies with $M_{\ast} = 4-8 \times 10^{10} \, \Msun$, within a factor of $\sim$2 from the value found in this work.

Despite the mathematical similarities in the merger rates of galaxies and DM haloes, it is interesting to note an important difference between their overall growth mechanisms. Since star formation `quenching' has no analogue in the case of DM haloes, the relative contributions from the two main channels of DM halo growth -- mergers and smooth accretion -- are independent of halo mass (approximately 60 and 40 per cent, respectively; e.g. \citealt{Genel2010}; \citealt{Fakhouri2010}). On the other hand, the importance of mergers for \textit{galaxy} growth changes substantially with stellar mass, from being statistically negligible in the case of small galaxies to becoming the dominant growth mechanism for galaxies with stellar masses above a few times $10^{11} \Msun$.

Finally, Fig. \ref{fig:macc_vs_mass} can be readily used to make predictions about the instantaneous rate of stellar mass growth by mergers and its importance relative to dissipative processes. For example, the left panel from Fig. \ref{fig:macc_vs_mass} shows that the stellar mass of a Milky Way-sized galaxy with $M_{\ast} \approx 6 \times 10^{10} \Msun$ \citep{McMillan2011} grows by $\sim$1 per cent per Gyr due to mergers (of any mass ratio) at $z \approx 0.1$, but at the same time grows by $\sim$10 per cent per Gyr due to in situ star formation, which is the dominating factor for galaxies of this mass.

\subsection{A fitting function}\label{subsec:macc_fitting_formula}

In Table \ref{tab:fitting_formula} we provide a fitting function for the specific stellar mass accretion rate of galaxies as a function of descendant mass $M_0$, merger mass ratio $\mu$, and redshift $z$, along with the corresponding best-fitting parameters. The fit was carried out by minimizing a chi-squared merit function in log-space using a Markov chain Monte Carlo (MCMC) algorithm.\footnote{http://dan.iel.fm/emcee/current/ \citep{Foreman-Mackey2013}.} The data points were generated by creating three-dimensional bins in $M_0$, $\mu$, and $1+z$ with bin widths equal to factors of 2.0, 1.2, and 1.1, respectively, and subsequently calculating the specific stellar mass accretion rate inside each bin. Only mergers with $M_{0} > 10^{8} \, \Msun$, $\mu > 1/1000$, and $z < 4$ were considered in the fit. The parameter uncertainties are defined as the range between the 16th and 84th percentiles of the marginal probability distributions produced by the MCMC algorithm, which are found to be approximately Gaussian. The best-fitting parameters yield a reduced chi-squared statistic of 1.21, which indicates that the fit is a good description of the data, while keeping the number of free parameters relatively low.

We point out that the specific stellar mass accretion rate of galaxies has a much stronger mass dependence than the specific DM accretion rate of DM haloes. In particular, \cite{Fakhouri2010} and \cite{Genel2010} determined that the specific growth rate of DM haloes is proportional to $M_0^{0.1}$ and $M_0^{0.15}$, respectively, independently of redshift (in this case $M_0$ denotes the descendant \textit{halo} mass), whereas the fitting function given in Table \ref{tab:fitting_formula} implies that the specific stellar mass accretion rate of galaxies at $z=0$ is proportional to $M_0^{\alpha_0} \approx M_0^{0.2}$ at low masses and to $M_0^{\alpha_0 + \delta_0} \approx M_0^{0.9}$ at high masses. This strong, positive trend with mass, along with the fact that the specific star formation rate is a decreasing function of stellar mass, implies that the `importance' of mergers with respect to galaxy growth increases rapidly with stellar mass, as we quantify in the next section.

\section{The ex situ stellar mass fraction}\label{sec:ex_situ_stellar_mass_fraction}

In the previous section we have presented predictions from the Illustris simulation on the \textit{instantaneous} stellar mass accretion rates of galaxies. In the present section we address the related question of how much stellar mass do galaxies assemble by stellar accretion \textit{across their lifetimes}. Specifically, in the following paragraphs we present results on the \textit{ex situ stellar mass fraction}, i.e. the fraction of the total stellar mass of a galaxy which is contributed by stars that formed in other galaxies and were subsequently accreted as a consequence of the hierarchical growth of structures. We analyse its dependence on stellar mass, redshift, and a selection of galaxy properties, as well as in terms of the different stellar `components' that originated from different kinds of accretion events.

\subsection{General trends}\label{subsec:general_trends}

\begin{figure}
\begin{center}
	\includegraphics[width=8.5cm]{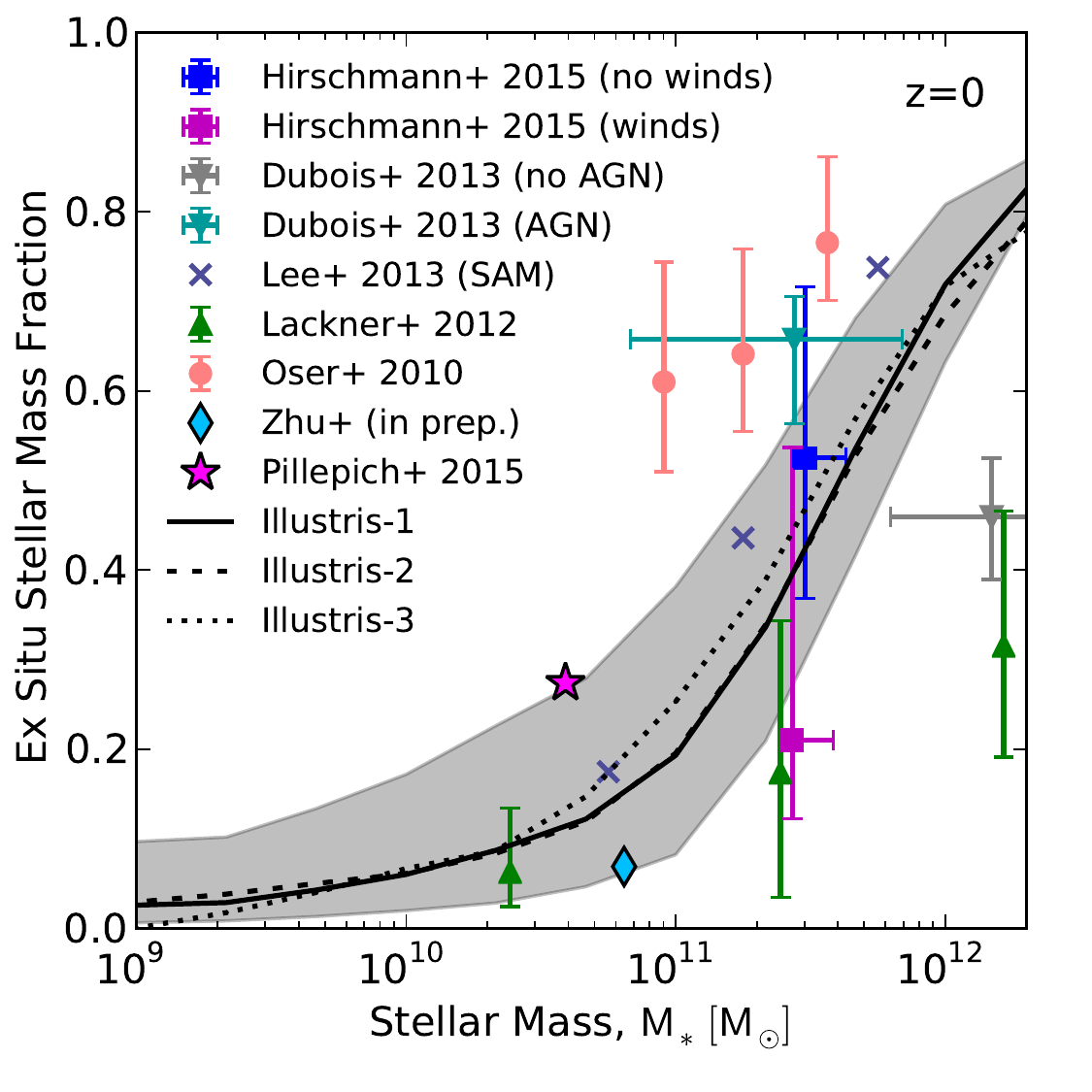}
	\caption{The fraction of accreted stellar mass (with respect to to the total) for galaxies at $z=0$, shown as a function of stellar mass $M_{\ast}$. The solid, dashed, and dotted black lines show the median results from the Illustris simulation carried out at different resolutions, with Illustris-1 being the highest resolution one. The grey shaded region represents the range between the 16th and 84th percentiles, or $1 \sigma$, while the different symbols show the ex situ stellar mass fraction as obtained in other theoretical models, both semi-analytic and hydrodynamic. The Illustris results here, as in the remainder of the paper unless otherwise stated, are obtained by adopting the stellar particle classification technique presented in Section \ref{subsec:stellar_particle_classification}. This figure shows that our determination of the ex situ stellar mass fraction is well-converged with resolution and also reduces the apparent tension among predictions from different theoretical models.}
	\label{fig:ex_situ_main}
\end{center}
\end{figure}

Fig. \ref{fig:ex_situ_main} shows the ex situ stellar mass fraction of galaxies at $z=0$ as a function of their stellar mass $M_{\ast}$, measured for the three different resolutions of the Illustris simulation: Illustris-1 (solid black), Illustris-2 (dashed black), and Illustris-3 (dotted black), of which Illustris-1 has the highest resolution. These three black curves represent median ex situ fractions for galaxies binned in $M_{\ast}$ and obtained with the stellar particle classification technique given in Section \ref{subsec:stellar_particle_classification}. The grey shaded region shows the corresponding $1\sigma$ scatter at a fixed stellar mass. 

The ex situ stellar mass fraction evidently has a strong dependence on stellar mass, ranging from $\sim$10 per cent for Milky Way-like galaxies to over 80 per cent for some of the most massive galaxies in the simulation. This trend is a consequence of both the increased rate of stellar mass accretion as well as the lower specific star formation rate displayed by more massive galaxies at later times, as summarized in Fig. \ref{fig:macc_vs_mass}. The magnitude of the scatter is significant, reaching a full width of $\sim$0.3 for galaxies with $M_{\ast} \approx 10^{11} \Msun$. For these galaxies the ex situ fraction can vary from 8 to about 38 per cent of their total stellar mass within the $1\sigma$ scatter, while for Milky Way-like galaxies the ex situ fraction ranges from 5 to 30 per cent of the total stellar mass. This is the result of the wide range of merging histories and halo formation times that are naturally produced by a cosmological simulation, as we will explicitly demonstrate in Section \ref{subsec:f_acc_by_type}.

The Illustris predictions at different resolutions are in excellent agreement with each other, which demonstrates that our classification scheme for ex situ particles, as well as the underlying physics that determines the stellar mass assembly of galaxies, are well-converged with numerical resolution. The different symbols in Fig. \ref{fig:ex_situ_main} show predictions from previous theoretical works, both from semi-analytic models of galaxy formation \citep{Lee2013} and from hydrodynamic zoom-in simulations \citep[][Zhu et al., in prep.]{Oser2010, Lackner2012a, Dubois2013, Pillepich2015, Hirschmann2015}. The error bars from \cite{Oser2010} and \cite{Lackner2012a} show interquartile ranges divided into three stellar mass bins. The data points from \cite{Hirschmann2015} indicate the median and $1\sigma$ scatter of the 10 galaxies considered in their study, showing results from their model both with strong galactic winds (magenta) and without them (blue). Similarly, the data points from \cite{Dubois2013} show the median and $1\sigma$ scatter of their 6 simulated galaxies, including results from their model both with AGN feedback (cyan) and without it (grey). We find that the ex situ stellar mass fraction in the Illustris simulation lies close to the `median' prediction from all the other studies, reducing the apparent tension among them.

We caution, however, that a direct comparison among the different studies shown in Fig. \ref{fig:ex_situ_main} is not a priori fully consistent, given the possibly different operational definitions adopted for the ex situ classification (e.g. including or excluding satellites) and for the galaxy stellar mass itself. For example, the results from \cite{Pillepich2015} are reported by excluding the contribution from satellites to the total ex situ fraction (i.e. by applying a correction to their nominal ex situ fraction based on information provided in the same paper), while we note that the results by \cite{Oser2010} and \cite{Dubois2013} are given as a function of the stellar mass measured within $0.1 R_{\rm vir}$ and are not corrected here. It is worthwhile to note that, if in Illustris we measured the ex situ fraction not across the whole \textsc{subfind} halo (as done throughout) but within a sphere of twice the stellar half-mass radius, the ex situ fractions would be lower at all masses, by about 5 per cent at the high-mass end and by $\sim$30--40 per cent for less massive galaxies.

\begin{figure*}
  \centerline{
	  \vbox{
			\hbox{
				\includegraphics[width=7.5cm]{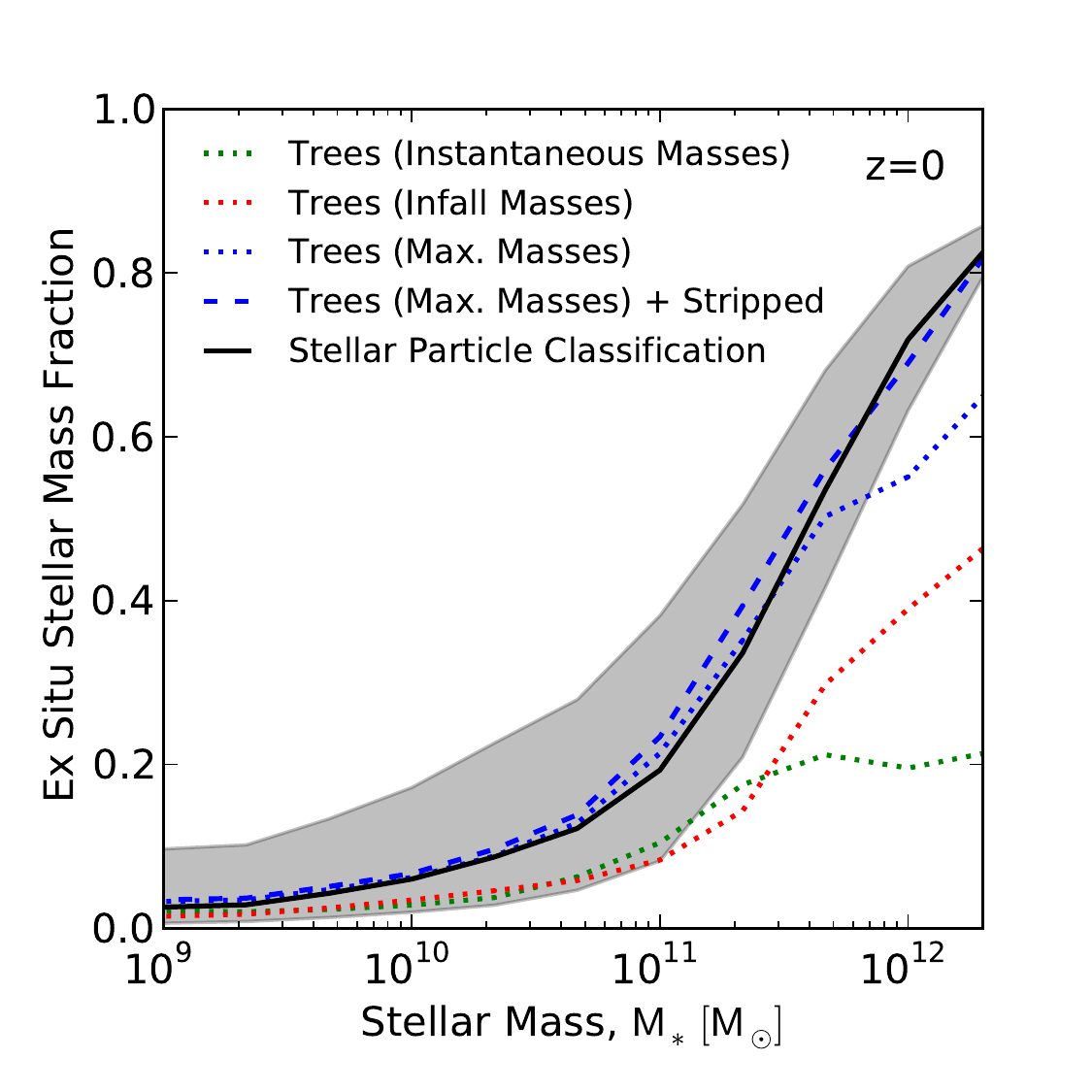}
				\includegraphics[width=7.5cm]{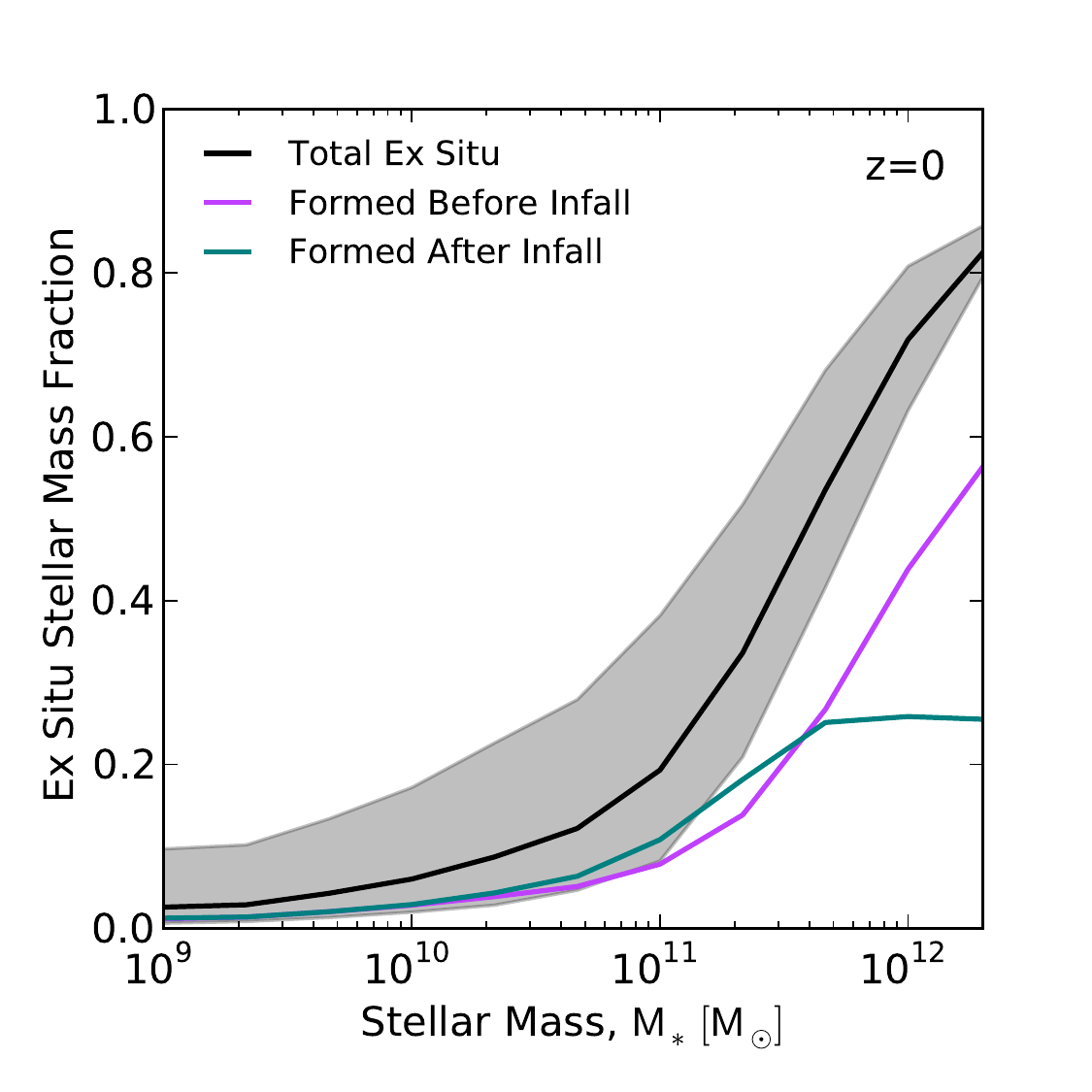}
			}
			\hbox{
				\includegraphics[width=7.5cm]{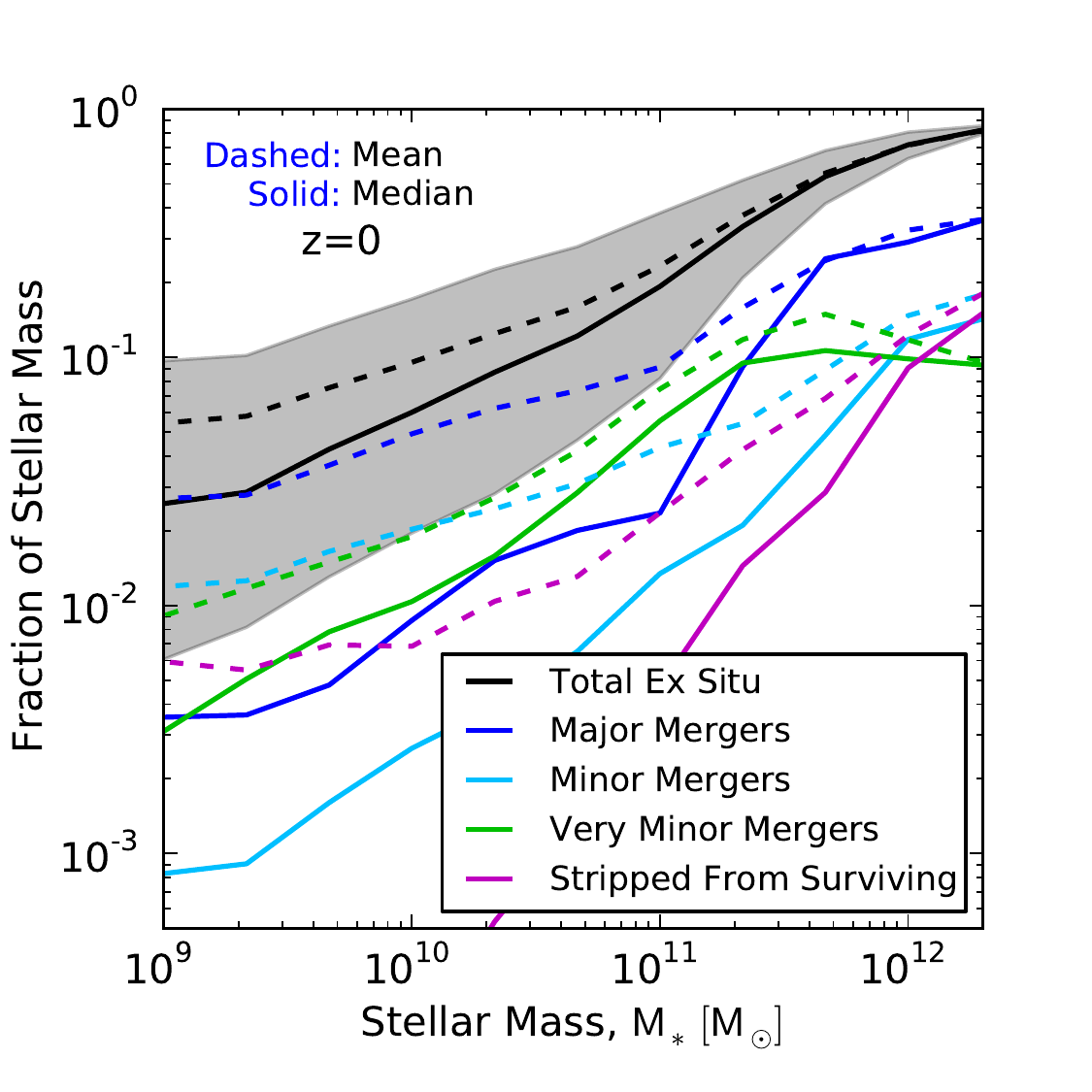}
        				\includegraphics[width=7.5cm]{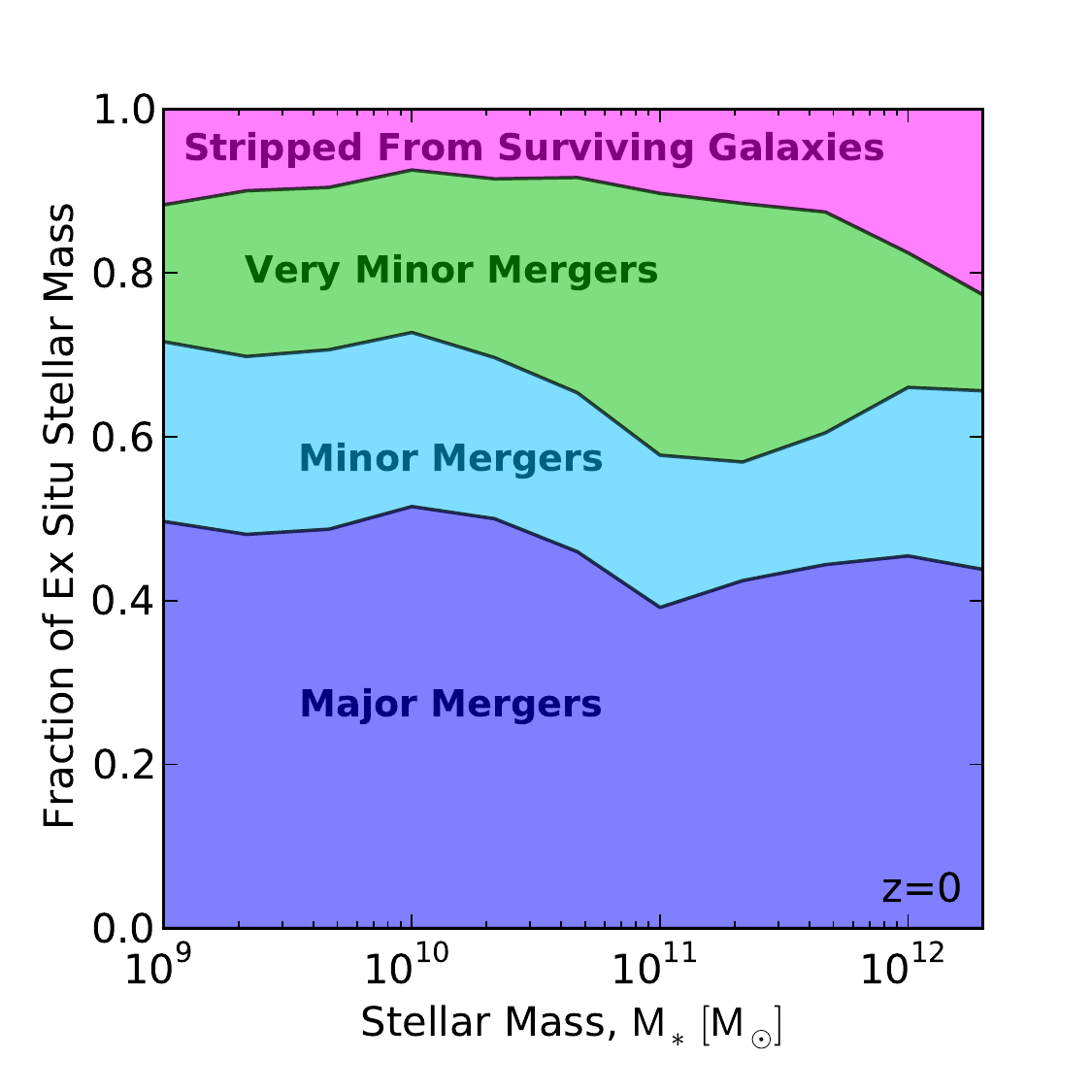}
			}	
		}
	}
	\caption{The ex situ stellar mass fraction of galaxies at $z=0$ as a function of stellar mass $M_{\ast}$, calculated with different methods or showing different particle selections. In all panels except the bottom-right one, the solid black curve is the median ex situ fraction from Fig. \ref{fig:ex_situ_main}, with the grey shaded region representing the range between the 16th and 84th percentiles. Top-left: the different lines show different methods of calculating the ex situ stellar mass fraction. All curves except the black one represent calculations carried out with the merger trees alone (instead of the stellar particle classification technique) by identifying all mergers and adding the masses of the secondary progenitors right before the merger (dotted green), at infall (dotted red), and at $t_{\rm max}$ (dotted blue). The latter gets close to the prediction from the stellar particle classification scheme, but is missing the contribution from stars that were stripped from surviving galaxies, which cannot be determined with the merger trees alone. If this missing component is added by adopting information from the stellar particle classification scheme (dashed blue), the two calculations become consistent with each other. On the other hand, the dotted red and dotted green lines both underestimate the ex situ stellar mass fraction, in the former case because stars formed after infall are not taken into account and in the latter because satellites have already lost a substantial amount of stellar mass due to stripping. Top-right: the two coloured lines show the amount of ex situ stellar mass that was formed before and after infall (as defined in Section \ref{subsec:stellar_particle_classification}). Bottom: The contributions to the ex situ stellar mass of galaxies at $z=0$ from stars with different accretion origins, shown as a function of stellar mass $M_{\ast}$. The different colours correspond to stars accreted in major mergers (blue, $\mu > 1/4$), minor mergers (cyan, $1/10 < \mu < 1/4$), very minor mergers (green, $\mu < 1/10$), or stripped from \textit{surviving} galaxies (magenta). The latter component includes stellar mass transferred during flyby events or during the first passages of an ongoing merger. Bottom-left: The dashed and solid lines correspond to mean and median quantities, respectively. Bottom-right: the \textit{average} fraction of ex situ stellar mass contributed by stars with different accretion origins (to be compared with the dashed lines from the left-hand panel).}
	\label{fig:ex_situ_allinone}
\end{figure*}

In Fig. \ref{fig:ex_situ_allinone} we further expand on various techniques to measure the ex situ fraction and on the origin of the accreted stars. The top-left panel of Fig. \ref{fig:ex_situ_allinone} compares different ways of calculating the ex situ stellar mass fraction: the solid black line, obtained with the stellar particle classification scheme described in Section \ref{subsec:stellar_particle_classification} and already shown in Fig. \ref{fig:ex_situ_main}, is compared to the dotted curves corresponding to calculations carried out with the merger trees alone. In the latter case, the ex situ stellar mass is obtained by finding all the mergers a given galaxy has ever had and adding up the stellar masses of the so-identified secondary progenitors. There is some freedom of choice regarding the stage of the merger at which the progenitor masses are measured: the dotted blue, dotted red, and dotted green lines show the effects from taking the progenitor masses at $t_{\rm max}$ (when the maximum stellar mass is reached), at infall\footnote{If the secondary progenitor itself undergoes additional mergers with other objects after infall, we also include the stellar masses at infall of such objects.} (right before the progenitor joins the same parent FoF group as the main progenitor), and right before the merger, respectively. 

In previous work \citep[][figure 5]{Rodriguez-Gomez2015} we showed that the maximum stellar mass of galaxies is usually reached \textit{after} infall. Indeed, for satellites merging with central galaxies at $z=0$, we found that $t_{\rm max}$ and $t_{\rm infall}$ typically happen 2--3 Gyr and $\sim$6 Gyr \textit{before} the time of the merger, respectively. In the same study we recommended measuring merger stellar mass ratios at $t_{\rm max}$, and indeed here we find this definition to be the one that gets closer to the more refined method of classifying stellar particles individually. However, even with this definition, the calculation is neglecting the contribution from stars that were stripped from surviving galaxies, which cannot be taken into account using the merger trees alone. However, if this missing component is included by adopting information from the stellar particle classification technique (dashed blue), the calculation becomes fully consistent with the more complete estimate given by the solid black line.

The other two definitions -- measuring progenitor masses at infall (dotted red) or right before the merger (dotted green) -- tend to underestimate the ex situ stellar mass fraction, although for different physical reasons. If the stellar mass of a progenitor is measured at infall, then the corresponding ex situ stellar mass cannot include the contribution from stars formed in galaxies \textit{after} infall. This results in an underestimate of the ex situ stellar mass by $\sim$50 per cent, with some galaxy-to-galaxy variation, suggesting that a significant amount of stellar mass can form in satellites also while they are orbiting the gravitational potential of a more massive companion. On the other hand, if the stellar mass of a progenitor is taken right before the moment of the final coalescence, the measurement cannot take into account the stellar material that has already been stripped from the progenitor, both tidally and numerically (e.g. due to insufficient density contrast), therefore also underestimating the fraction of stars accreted from other galaxies.

The top-right panel of Fig. \ref{fig:ex_situ_allinone} further illustrates in physical terms the reasons why the different methods presented in the top-left panel give different estimates of the accreted stellar fraction. Here we show the amount of ex situ stellar mass that was formed \textit{before infall} and \textit{after infall}, as defined in Section \ref{subsec:stellar_particle_classification}. About 60 per cent of the ex situ stellar mass in $M_{\ast} \approx 10^{10.5} \, \Msun$ galaxies was formed after infall, i.e. in satellite galaxies orbiting Milky Way-like hosts, but this fraction becomes smaller as the mass of the host increases, most likely as a consequence of the stronger environmental effects exerted by more massive hosts (see \citealt{Sales2014a}, for a discussion). We note that the `before infall' component from the top-right panel is closely related to the dotted red line from the top-left panel, the latter of which is an approximation of the ex situ fraction obtained by adding the infall masses of all progenitors recursively (as mentioned above). The small differences between the two calculations can be attributed to the fact that the merger-tree-only estimate cannot take into account the `mass transfer' that results from stripping during galaxy interactions which are not recorded as mergers.

Finally, as mentioned in Section \ref{subsec:stellar_particle_classification}, for each ex situ stellar particle we are able to determine the stellar mass ratio of the merger in which it was accreted. This allows us to evaluate the relative contributions from major ($\mu > 1/4$), minor ($1/10 < \mu < 1/4$), and very minor ($\mu < 1/10$) mergers to the final stellar mass of a galaxy, as well as the contribution from stars in the \textit{stripped from surviving galaxies} category -- namely, stars that have been stripped from a galaxy which is either in the process of merging with the descendant host or is undergoing a flyby passage. The contributions from these different components of the ex situ stellar mass at $z=0$ are plotted as a function of stellar mass in the bottom row of Fig. \ref{fig:ex_situ_allinone}. The blue, cyan, and green colours represent major, minor, and very minor mergers, while magenta corresponds to stars that were stripped from surviving galaxies, and black shows the total fraction of ex situ stellar mass.

The bottom-left panel shows that the mean (dashed) and median (solid) stellar mass fractions can be very different from each other. In general, given the discrete nature of mergers, the mean is larger than the median. This happens because very few galaxies at low masses have an important contribution from major or minor mergers, but the few of them that have recently undergone such mergers tend to have significantly higher ex situ fractions, raising the value of the mean. The two measures converge as we consider galaxies of higher stellar masses, whose growth is dominated by accretion.

The bottom-right panel of Fig. \ref{fig:ex_situ_allinone} shows the \textit{average} (hence the mean) contributions to the ex situ stellar mass from the ex situ components shown on the left-hand panel. These fractions are approximately constant with respect to stellar mass, with their trends and values being consistent with the results from Section \ref{sec:stellar_mass_accretion}. It is worth mentioning that the near mass-independence of the relative contributions to the \textit{galaxy merger rate} from mergers with different mass ratios \citep[][figure 7]{Rodriguez-Gomez2015} leads to a specific stellar mass accretion rate with the same property (Fig. \ref{fig:macc_vs_mass}), which in turn results in an ex situ stellar mass fraction at $z=0$ for which the same statement is approximately valid. Consequently, the blue, cyan and green lines used in \cite{Rodriguez-Gomez2015} as well as in this work, where we examine more complex quantities than the galaxy merger rate, always appear to be roughly parallel to each other.

\begin{figure*}
	\includegraphics[width=16cm]{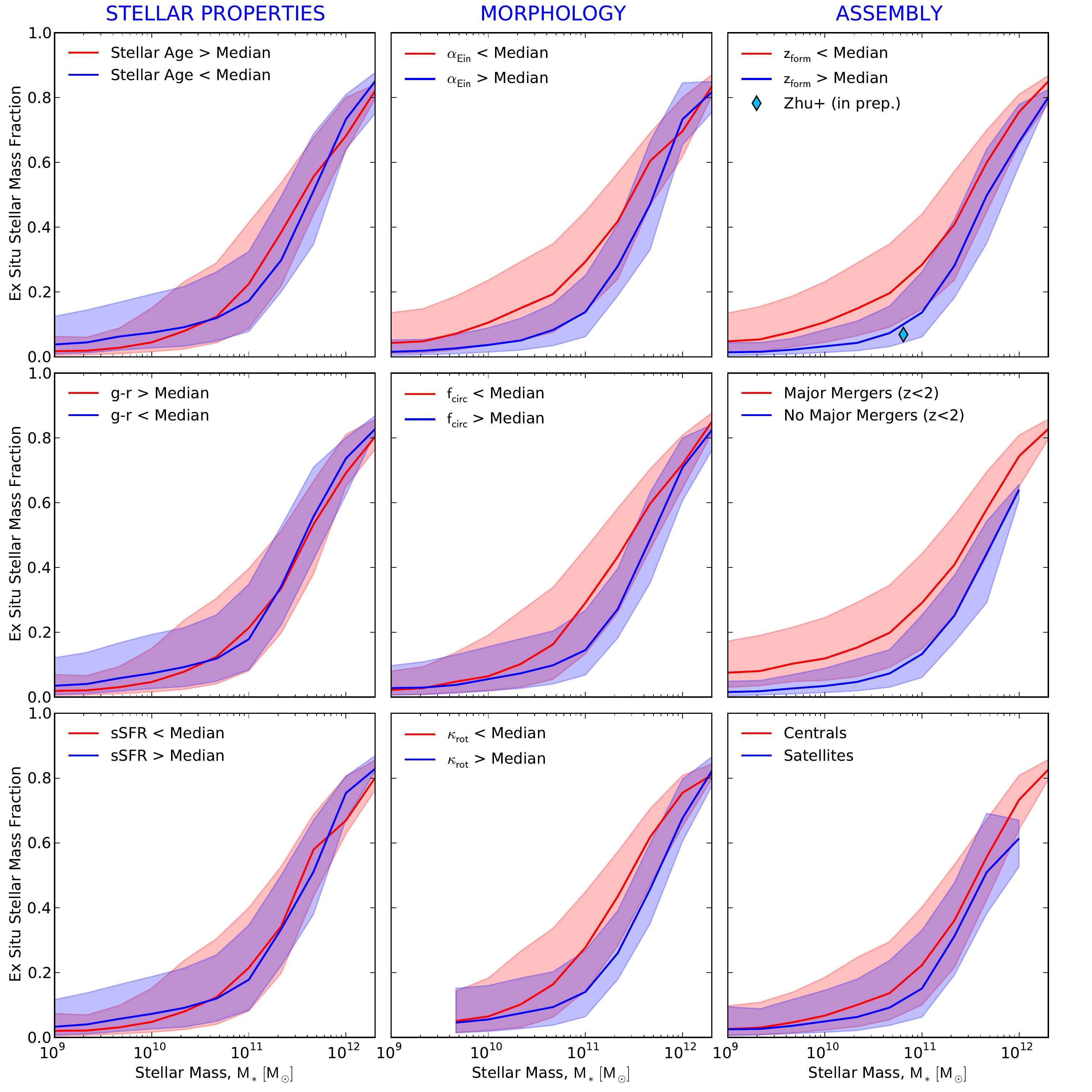}
	\caption{The median ex situ stellar mass fraction as a function of stellar mass $M_{\ast}$, calculated for galaxies at $z=0$. The red and blue lines show the correlation (at a fixed stellar mass) with respect to various galaxy properties, which are divided into three broad categories: stellar properties (left), morphology (middle), and halo assembly (right). These galaxy properties are described in detail in Section \ref{subsec:f_acc_by_type}. The shaded regions indicate the 16th to 84th percentile ranges. The blue diamond in the upper right panel corresponds to the Milky Way-like galaxy studied in Zhu et al. (in prep.), which has a relatively early formation time.}
	\label{fig:f_acc_vs_mass_by_type}
\end{figure*}

\begin{figure*}
\centerline{
	\hbox{
	\includegraphics[width=7.5cm]{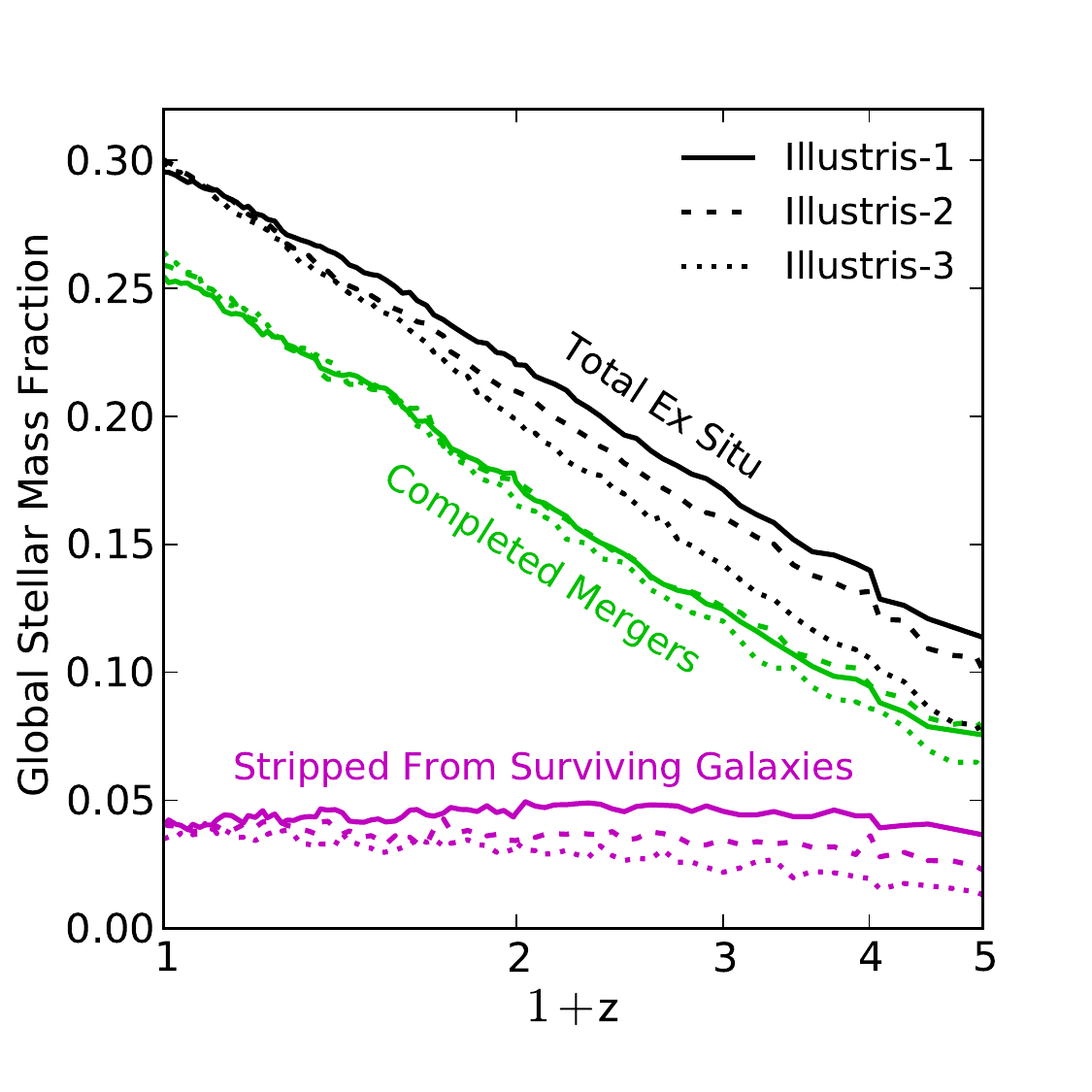}
  \includegraphics[width=7.5cm]{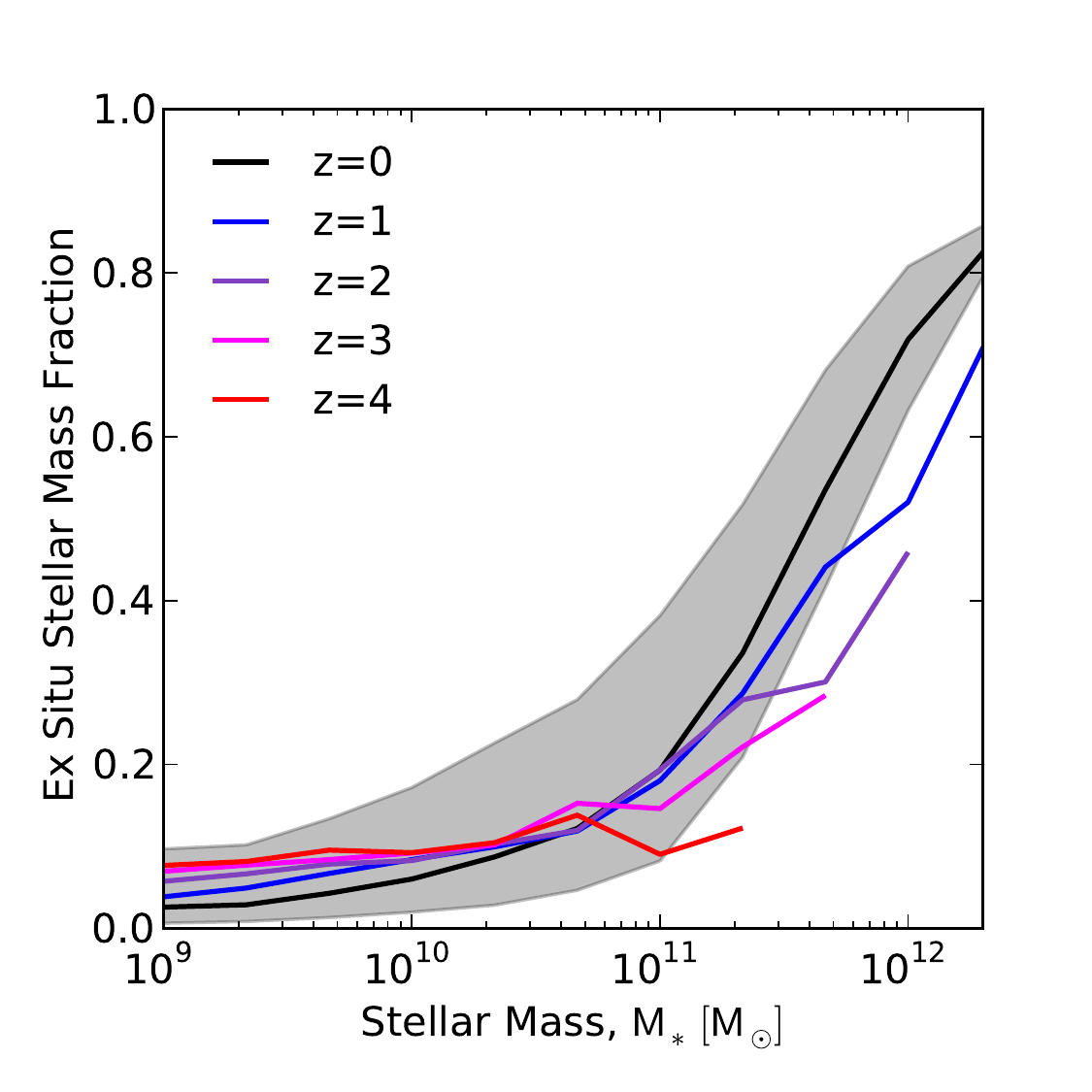}
	}}
	\caption{Redshift evolution of the ex situ stellar mass fraction. Left: the fraction of ex situ stars across the whole Illustris volume as a function of redshift, obtained by summing over all galaxy stellar masses (black lines). The green and magenta curves denote the contributions from stars that were accreted via completed mergers (of any mass ratio) and from those that were stripped from surviving galaxies, respectively. The solid, dashed and dotted lines correspond to the Illustris-1, 2, and 3 simulations, of which Illustris-1 has the highest resolution. Right: the ex situ stellar mass fraction as a function of galaxy stellar mass, shown at different redshifts. The black line corresponds to $z=0$ and is shown for reference (it is identical to the solid black line in Figs. \ref{fig:ex_situ_main} and \ref{fig:ex_situ_allinone}), while the coloured lines show the ex situ stellar mass fraction at $z =$ 1, 2, 3, and 4.}
	\label{fig:f_acc_vs_z_snaps}
\end{figure*}

\subsection{Dependence on galaxy type and halo assembly history}\label{subsec:f_acc_by_type}

We have just shown in Section \ref{subsec:general_trends} that the ex situ stellar mass fraction has a very large scatter at a fixed stellar mass. In the current section we investigate the origin of such scatter by considering all galaxies at $z=0$ with stellar masses $> 10^{9} \, \Msun$ and separating them -- at a fixed stellar mass -- according to different quantities, which can be grouped into three broad categories: stellar properties, morphology, and halo assembly.

Fig. \ref{fig:f_acc_vs_mass_by_type} shows the ex situ stellar mass fraction as a function of stellar mass, showing two components (red and blue) corresponding to galaxy populations separated by the criteria indicated in the labels. In particular, when separating galaxies by the median value of a given galaxy property (e.g. stellar age), the median is calculated within each mass bin, rather than for the full galaxy population. The shaded regions show the 16th to 84th percentile ranges. The left, centre, and right panels correspond to galaxies separated by stellar properties, morphology, and assembly history, respectively. Below is a brief description of the quantities used to classify galaxies, along with some observations about the implications from Fig. \ref{fig:f_acc_vs_mass_by_type}.

\subsubsection{Stellar properties}

We consider the correlation between the ex situ stellar mass fraction and galactic stellar properties (at a fixed stellar mass) by measuring the following quantities:

\begin{itemize}
  \item \textit{Stellar age:} The stellar mass-weighted median age of all the stellar particles in the subhalo.\\
  
  \item \textit{g-r:} A measure of galaxy colour. The \textit{g} and \textit{r} magnitudes are based on the sum of the luminosities of all the stellar particles in the subhalo \citep[see][for more details]{Nelson2015}.\\
  
  \item \textit{sSFR:} The specific star formation rate, defined as the star formation rate of all the gas cells in the subhalo, divided by the total stellar mass.\\
\end{itemize}

In general, once the galaxy stellar mass is fixed, the ex situ stellar mass fraction shows a very weak, if not null, secondary dependence on the stellar properties described above. We find the same result when separating galaxies by their cold gas fraction (not shown).

\subsubsection{Morphology}

We consider the correlation of the ex situ stellar mass fraction with galactic morphology by using the following calculated parameters:

\begin{itemize}
  \item $\alpha_{\rm Ein}$: The `$\alpha$' exponent of an Einasto profile fitted to the \textit{stellar} content of each subhalo. The fit was carried out by maximizing the log-likelihood that the stellar particles are drawn from a spherically-symmetric probability distribution proportional to an Einasto profile, which is given by
  \begin{equation}
    \rho(r) = \rho_{-2} \exp\left(\frac{-2}{\alpha}\left[\left(\frac{r}{r_{-2}}\right)^{\alpha} - 1\right] \right),
  \end{equation}
where $r_{-2}$ is the radius at which the logarithmic slope of the profile becomes $-2$, and $\rho_{-2}$ is the density at $r = r_{-2}$. Only stellar particles within twice the stellar half-mass radius were considered for the fit. An Einasto profile is mathematically similar to a S\'ersic profile, which is typically used to fit observations of surface brightness profiles. However, we point out that $\alpha_{\rm Ein}$ is inversely proportional to the S\'ersic index $n$, so that $\mathrm{\alpha_{Ein} < \, Median}$ ($\mathrm{\alpha_{Ein} > \, Median}$) corresponds to early-type (late-type) galaxies. \\

  \item $f_{\rm circ}$: The fraction of `disc' stars in the subhalo, defined as those with $\epsilon > 0.7$, where the `orbital circularity' parameter $\epsilon = J_{z} / J(E)$ of a star is the ratio between its specific angular momentum $J_z$ and the maximum specific angular momentum possible at the specific binding energy $E$ \citep{Abadi2003, Marinacci2013}. The $z$-direction corresponds to the total angular momentum of the stars in the subhalo. All stellar particles within 10 times the stellar half-mass radius were considered \citep[see][table C2, for more details]{Nelson2015}. \cite{Genel2015} found the scaling relations of Illustris galaxies would remain approximately unchanged if only particles within 5 times the stellar half-mass radius were considered instead. \\
  
  \item $\kappa_{\rm rot}$: The fraction of kinetic energy that is invested in ordered rotation, defined in \cite{Sales2012} as 
  \begin{equation}
    \kappa_{\rm rot} = \frac{K_{\rm rot}}{K} = \frac{1}{K}\sum\frac{1}{2}\left(\frac{J_z}{R}\right)^{2}.
  \end{equation}
  In this case the $z$-direction coincides with the angular momentum of the stellar content of the subhalo. The sum was carried out over all stellar particles within twice the stellar half-mass radius.
\end{itemize}

The panels from the middle column of Fig. \ref{fig:f_acc_vs_mass_by_type} show that the ex situ stellar mass fraction is correlated with morphology. At a fixed stellar mass, spheroidal galaxies (red lines) have a higher accreted fraction relative to their disc-like counterparts (shown in blue), although the scatter is large. Observationally, this result is supported by \cite{DSouza2014}, who found that galaxies with high concentration have higher fractions of accreted material than their low-concentration counterparts (at a fixed stellar mass). We note, however, that when using kinematic morphologies (middle and bottom panels) this separation becomes very weak for Milky Way-sized galaxies, and in fact completely disappears at $M_{\ast} \sim 10^{10} \, \Msun$, in good agreement with previous results \citep{Sales2012}. The significant overlap in the fraction of accreted stars for disc- and spheroid-dominated galaxies suggests that mergers alone cannot be the only driver of galaxy morphology.

\subsubsection{Assembly}

\begin{figure*}
	\includegraphics[width=17.5cm]{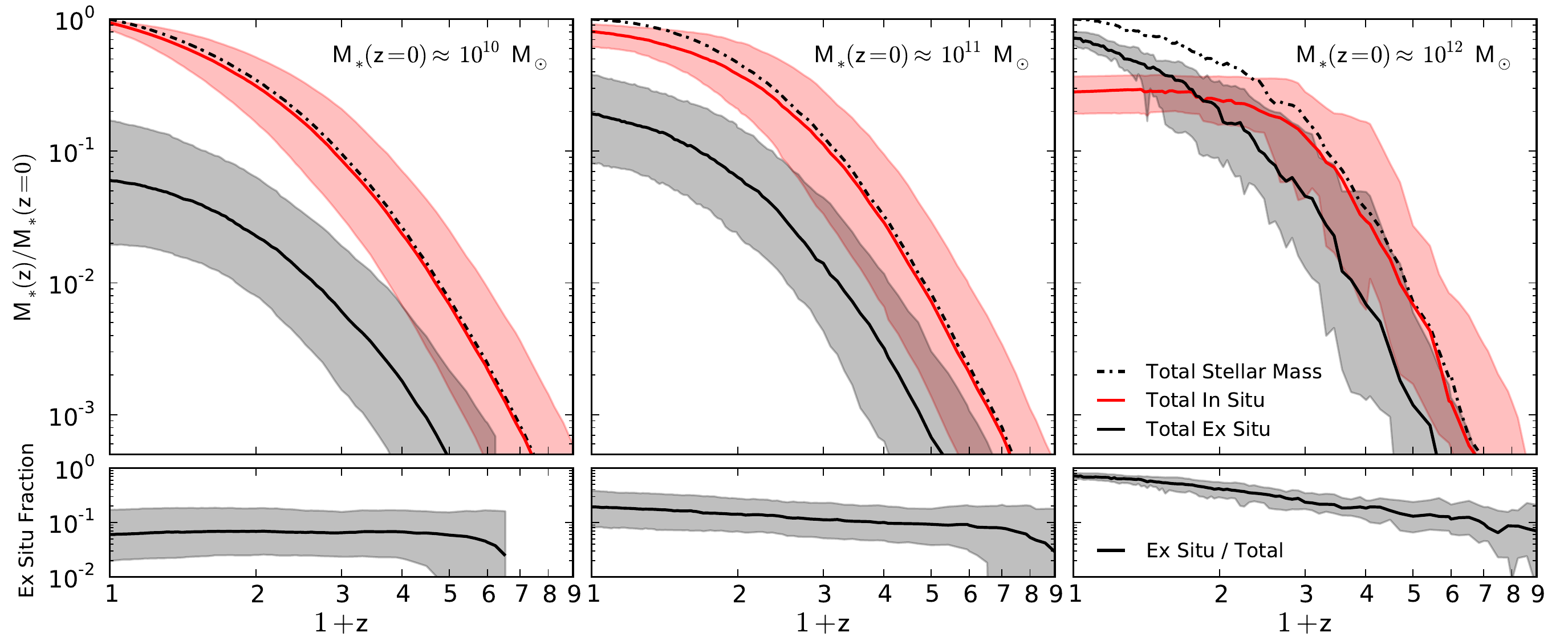}
	\caption{The median stellar mass history of galaxies selected at $z=0$ and tracked back in time using the merger trees. The left, centre, and right panels correspond to $z=0$ stellar mass bins centred at $10^{10}$, $10^{11}$ and $10^{12} \, \Msun$, respectively (each mass bin is a factor of $\sim$2 wide). Each individual stellar mass has been normalized by the stellar mass of the corresponding descendant at $z=0$. The dot-dashed black line shows the evolution of the total stellar mass, while the solid red and solid black lines show the contributions from stars that were formed in situ and ex situ, respectively. The shaded regions indicate the 16th to 84th percentile ranges. The bottom panels show the evolution of the median `instantaneous' ex situ stellar mass fraction, which corresponds to the ratio between the solid and dot-dashed black lines from the upper panels. We note that the so-called `two-phase' picture of galaxy formation only happens for the galaxies in our most massive bin.}
	\label{fig:f_acc_vs_z}
\end{figure*}

Finally, we characterize the assembly of the underlying DM haloes with the following information:

\begin{itemize}
  \item $z_{\rm form}$: The redshift at which the \textit{total mass} of the subhalo in question reached \textit{half of its maximum value}. The maximum total mass is not necessarily attained at $z=0$, especially in the case of satellites. In order to smooth out short-term noise fluctuations in the mass of the subhalo, we fit a 7th order polynomial to the total mass history of the subhalo. An alternative method consists in convolving the mass history with a median box filter of full width equivalent to 5 snapshots \citep{Bray2016}. We find that these methods are in good agreement with each other. We caution that $z_{\rm form}$ is referred to as \textit{halo formation time} or \textit{halo formation redshift}, even though the calculation is carried out for the corresponding subhalo (i.e. \textsc{subfind} halo) of each galaxy. \\
  
  \item \textit{Major mergers since $z=2$:} Mergers are identified as described in \cite{Rodriguez-Gomez2015}: the stellar mass ratio of each merger is taken at $t_{\rm max}$ (the time when the secondary progenitor reaches its maximum stellar mass) and it is checked that each merger has a well-defined `infall' moment. The merger itself is considered to take place at the time of the final coalescence.\\
  
  \item \textit{Central/satellite status:} As determined by the \textsc{subfind} algorithm (see Section \ref{subsec:illustris} for more details).
\end{itemize}

The ex situ stellar mass fraction shows a strong correlation with halo formation time (parametrized by $z_{\rm form}$) and merging history (given by the number of major mergers since $z=2$). This means that galaxies inside haloes that assembled late and galaxies with violent merging histories have relatively higher ex situ stellar mass fractions. There is also a noticeable trend with respect to central/satellite status, but the differences are smaller because not all satellites have been equally subjected to the environmental effects, especially those which entered their parent halo at later times.

An interesting special case is the Milky Way-like galaxy described in \cite{Marinacci2013}, which is shown in Zhu et al. (in prep.) to have a relatively low ex situ stellar mass fraction, around 7 per cent (upper right panel from Fig. \ref{fig:f_acc_vs_mass_by_type}). Their simulation uses the same hydrodynamic code and physical model as those used in Illustris. Additionally, the merger trees and the stellar particle classification scheme employed by them are identical to the ones used in this work. Therefore, the low ex situ stellar mass fraction found in their study can only be attributed to scatter: consistently with the findings presented so far, the simulated halo from \cite{Marinacci2013} has a relatively quiet merging history and is also found to have assembled very early on, compared to other Illustris galaxies of similar mass (Zhu et al., in prep.).

Finally, we note that we also examined the dependence of the ex situ stellar mass fraction on galaxy overdensity (not shown), using a definition based on the distance to the 5th nearest neighbour with an $r$-band magnitude brighter than $-19.5$ \citep{Vogelsberger2014a}, but found no significant correlation. This null result is in agreement with \cite{Lackner2012a}, who compared the ex situ stellar mass fractions of galaxies in void and cluster environments, and also found no significant difference between the two populations.

\subsection{Redshift evolution}\label{subsec:redshift_evolution}

\begin{figure*}
	\includegraphics[width=18cm]{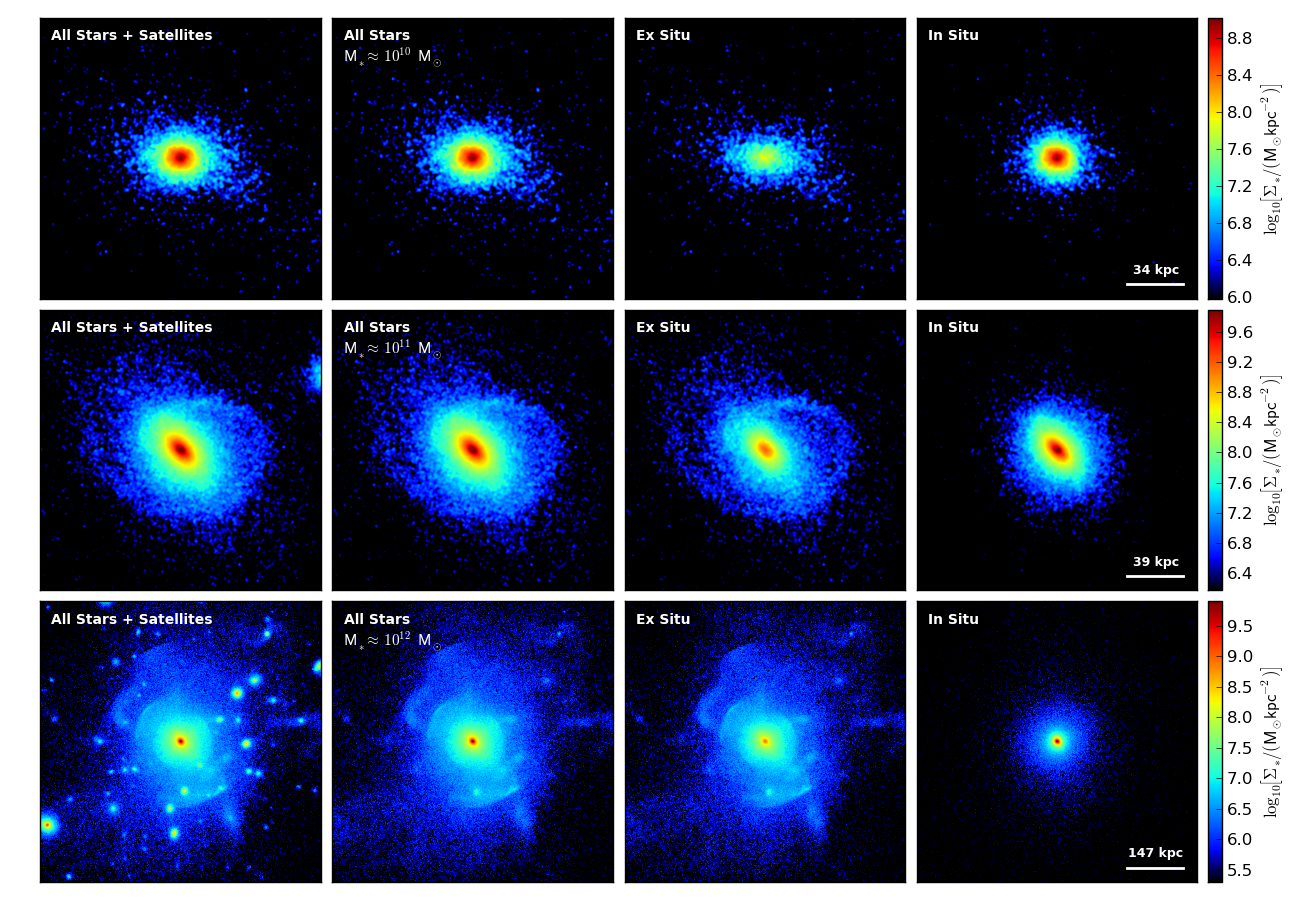}
	\caption{Projected stellar density maps at $z=0$. The top, middle, and bottom panels show randomly selected galaxies with stellar masses of approximately $10^{10}$, $10^{11}$, and $10^{12} \, \Msun$, respectively. From left to right, the different panels show (1) the total stellar content \textit{plus} satellites, if any (only applicable to centrals), (2) the total stellar content \textit{without} satellites (i.e. all stars that are bound only to the galaxy in question), (3) the ex situ stars, and (4) the in situ stars. Each image is a square with 20 stellar half-mass radii on a side, while the scale on the lower-right of the right hand panels corresponds to 4 stellar half-mass radii.}
	\label{fig:projected_density}
\end{figure*}

We find that the fraction of the total stellar mass that is classified as ex situ in the whole Illustris volume increases monotonically with time, reaching 17 per cent at $z=2$ and 30 per cent at $z=0$. This means that the majority of the stars found in galaxies today have been formed in situ, and this holds true at all available redshifts. We show this in the left-hand panel of Fig. \ref{fig:f_acc_vs_z_snaps}, where the black curves denote the \textit{global} ex situ fraction (obtained by summing the contributions from all galaxies) as a function of redshift. The green curves show the contribution from the subset of ex situ stars which were accreted via completed mergers (of any mass ratio), while the magenta curves indicate the contribution from stars that were stripped from surviving galaxies. The solid, dashed, and dotted lines represent the three different resolutions of Illustris, of which Illustris-1 is the highest. While the lower-resolution runs are well consistent with the Illustris-1 results at $z=0$ (in agreement with Fig. \ref{fig:ex_situ_main}), they underperform at higher redshifts when mergers are more frequent, since the lack of resolution limits the time during which two (or more) objects that are undergoing a merger can be resolved as separate entities, underestimating the number of surviving galaxies at any given time.

When considering the redshift evolution of the ex situ stellar mass fraction as a function of mass (right-hand panel of Fig. \ref{fig:f_acc_vs_z_snaps}), we find that the ex situ stellar mass fraction is an increasing function of cosmic time for galaxies above $M_{\ast} \approx 10^{11} \, \Msun$ but \textit{decreases} with time for galaxies below $M_{\ast} \approx 10^{10} \, \Msun$, while galaxies in the intermediate range ($10^{10}$--$10^{11} \, \Msun$) show little evolution with redshift. As we have seen in Fig. \ref{fig:macc_vs_mass}, these opposing trends can be understood by a cessation of in situ star formation at low redshifts in the most massive galaxies, and a decrease in the specific mass growth by accretion at low redshifts for dwarf galaxies. We note, however, that these redshift trends cannot be applied to individual galaxies, but only to galaxy populations selected at a constant stellar mass across different redshifts \citep[see][for an observationally motivated discussion on matching galaxy populations across different epochs]{Torrey2015}. In the following paragraphs we investigate the redshift evolution of the ex situ stellar mass fraction for individual galaxies (rather than temporally `disconnected' galaxy populations) by following them in time using the merger trees.

The upper panels from Fig. \ref{fig:f_acc_vs_z} show the (normalized) stellar mass histories of galaxies selected at $z=0$ and tracked back in time using the merger trees, showing the relative contributions of stars that were formed in situ (red) and ex situ (black), along with the evolution of the total stellar mass (dot-dashed black). The panels from left to right correspond to galaxies with $z=0$ stellar masses of approximately $10^{10}$, $10^{11}$ and $10^{12} \, \Msun$, respectively (each mass bin is a factor of $\sim$2 wide). The lower panels from Fig. \ref{fig:f_acc_vs_z} show that the `instantaneous' fraction of ex situ stellar mass (given by the ratio between the solid and dot-dashed black lines from the upper panels) displays different redshift evolution trends for galaxies of different masses. For galaxies in the low-mass bin, the ex situ fraction is approximately constant with redshift, while for galaxies in the medium-mass (high-mass) bin the ex situ fraction is found to mildly (strongly) increase with cosmic time.

As pointed out before, the stellar content of most galaxies is usually dominated by in situ star formation. However, for the galaxies in our most massive bin (Fig. \ref{fig:f_acc_vs_z}, right) there is a clear `crossover' point at $z \lesssim 1$, where the in situ and ex situ stellar components become comparable. We note that although this crossover between the two components has been presented before \citep[][figure 8]{Oser2010}, we find that it only happens for galaxies in our most massive bin, whereas Oser et al. observed a transition for all of the galaxies they studied, with halo masses ranging from $7.0 \times 10^{11} h^{-1}\Msun$ to $2.7 \times 10^{13} h^{-1}\Msun$ (stellar masses between $4 \times 10^{10} h^{-1}\Msun$ and $4 \times 10^{11} h^{-1}\Msun$). This implies that the `two-phase' picture of galaxy formation put forward by \cite{Oser2010} is only a good description for the most massive galaxies in our simulation, which reach stellar masses of $\sim 10^{12} \, \Msun$ at $z=0$. These differences can be partially attributed to the star formation and feedback prescriptions used by Oser et al., which resulted in very efficient star formation at early times and a large frequency of spheroidal systems with old stellar populations at later times.

The increased importance of in situ star formation relative to stellar accretion that we have found for most galaxies in our sample is in qualitative agreement with results from semi-empirical methods \citep[][figure 8]{Moster2012a}, as well as with hydrodynamic simulations of individual disc galaxies \citep[e.g.][]{Pillepich2015}. Furthermore, our results are in broad agreement with recent work by \cite{Vulcani2015}, who used observations of ultra-massive galaxies between $z=0.2$ and $z=2$ to study the growth of galaxies with $z=0$ descendants with $M_{\ast} > 10^{11.8} \, \Msun$, which roughly corresponds to our most massive bin ($M_{\ast} \approx 10^{12} \, \Msun$). Vulcani et al. found that the growth of massive galaxies is dominated by in situ star formation at $z \sim 2$, both star formation and mergers at $1 < z < 2$, and mergers alone at $z < 1$. We caution, however, that such observational estimates of galaxy growth are sensitive to the assumptions used to link progenitors and descendants across cosmic time, which is a non-trivial task \citep{Behroozi2013, Torrey2015}.

\begin{figure*}
	\centerline{
		\vbox{
		        \includegraphics[width=17.5cm]{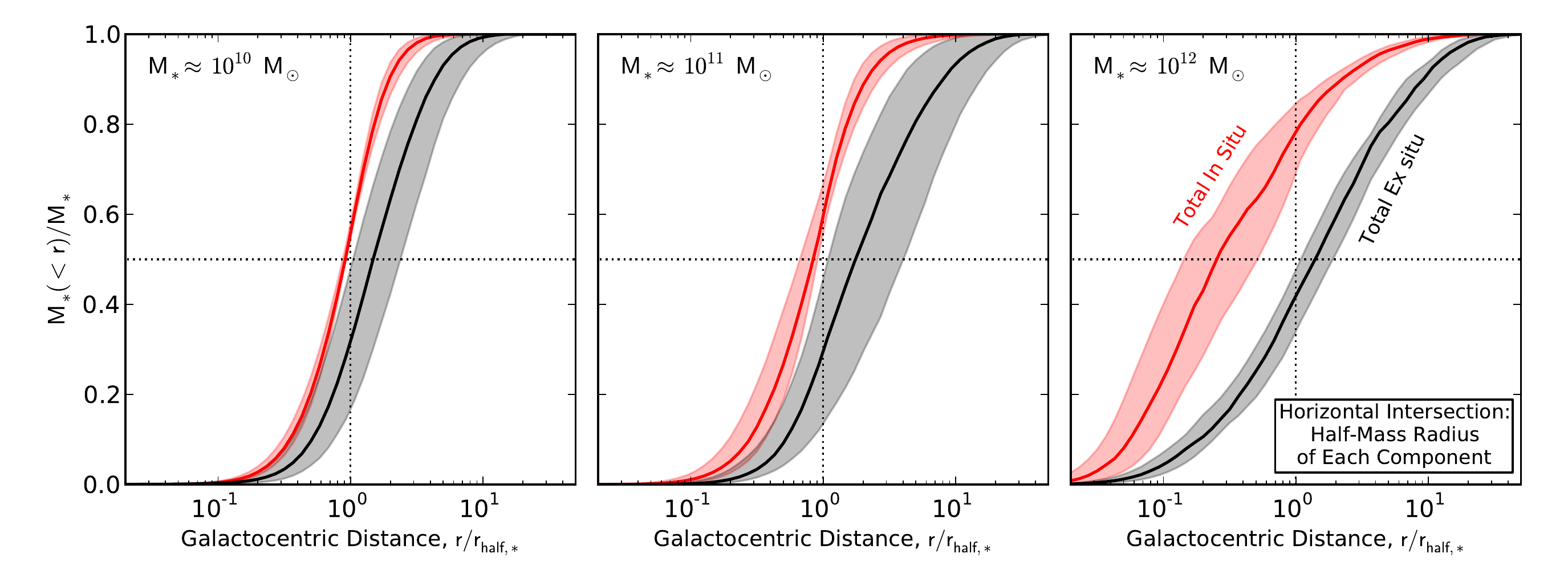}
			\includegraphics[width=17.5cm]{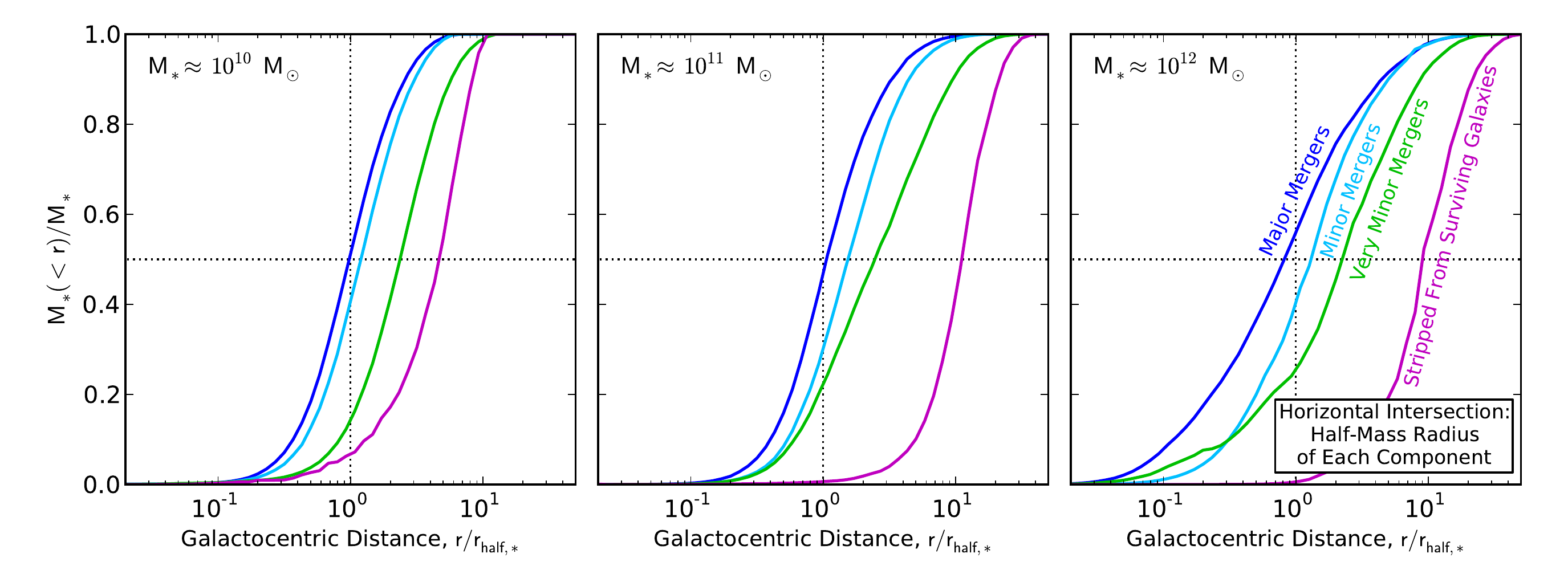}
			}
	}
\caption{Median \textit{cumulative} mass profiles (spherically averaged) of the different stellar components of galaxies. Each profile has been normalized to the total mass of the corresponding component, so that the intersections with the horizontal dotted line indicate the half-mass radii of the different components. The left, centre, and right panels correspond to the stellar mass bins centred at $10^{10}$, $10^{11}$ and $10^{12} \, \Msun$, respectively. Top: showing only the total in situ (red) and total ex situ (black) components, with the shaded regions indicating the 16th to 84th percentile ranges. Bottom: the same, but showing the different accretion origins of ex situ stars. This figure shows that, in general, galaxies display spatial segregation: stars that were formed in situ are found closest to the galactic centre, followed by stars accreted in major mergers, minor mergers, very minor mergers, and finally stars that were stripped from surviving galaxies.}
\label{fig:f_acc_vs_radius_cumulative}
\end{figure*}

\begin{figure*}
	\centerline{
		\vbox{
			\includegraphics[width=17.5cm]{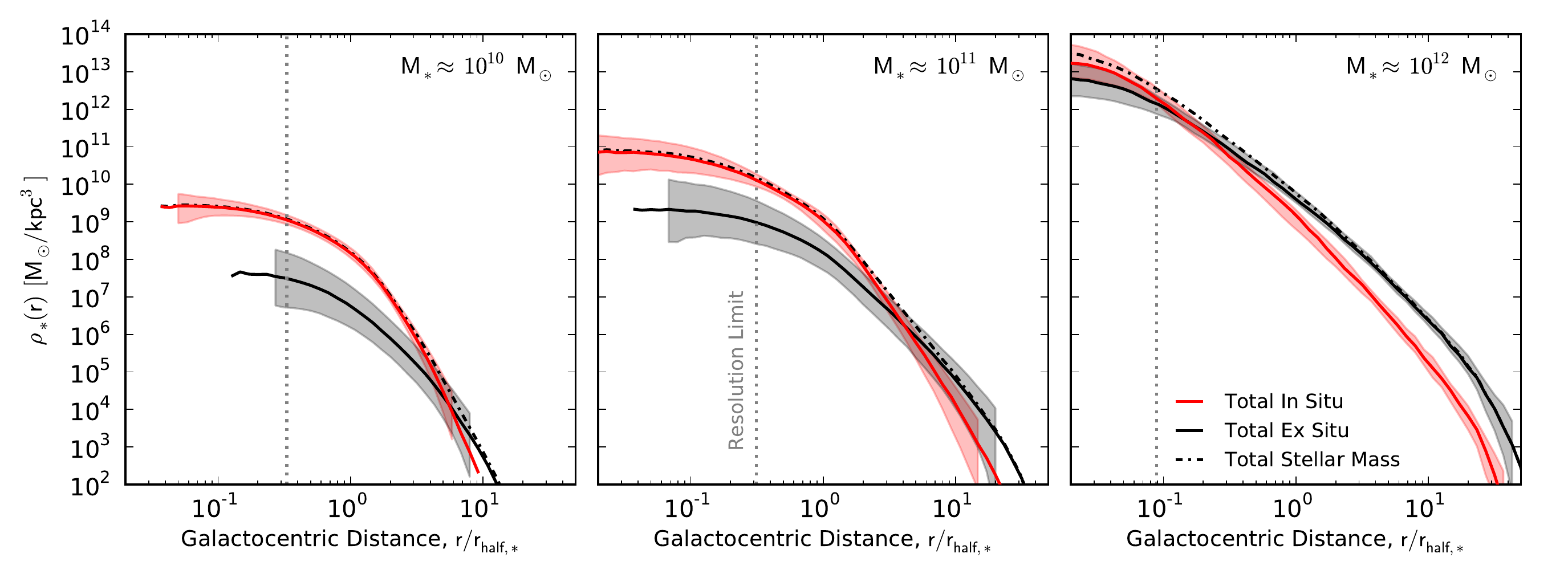}
			\includegraphics[width=17.5cm]{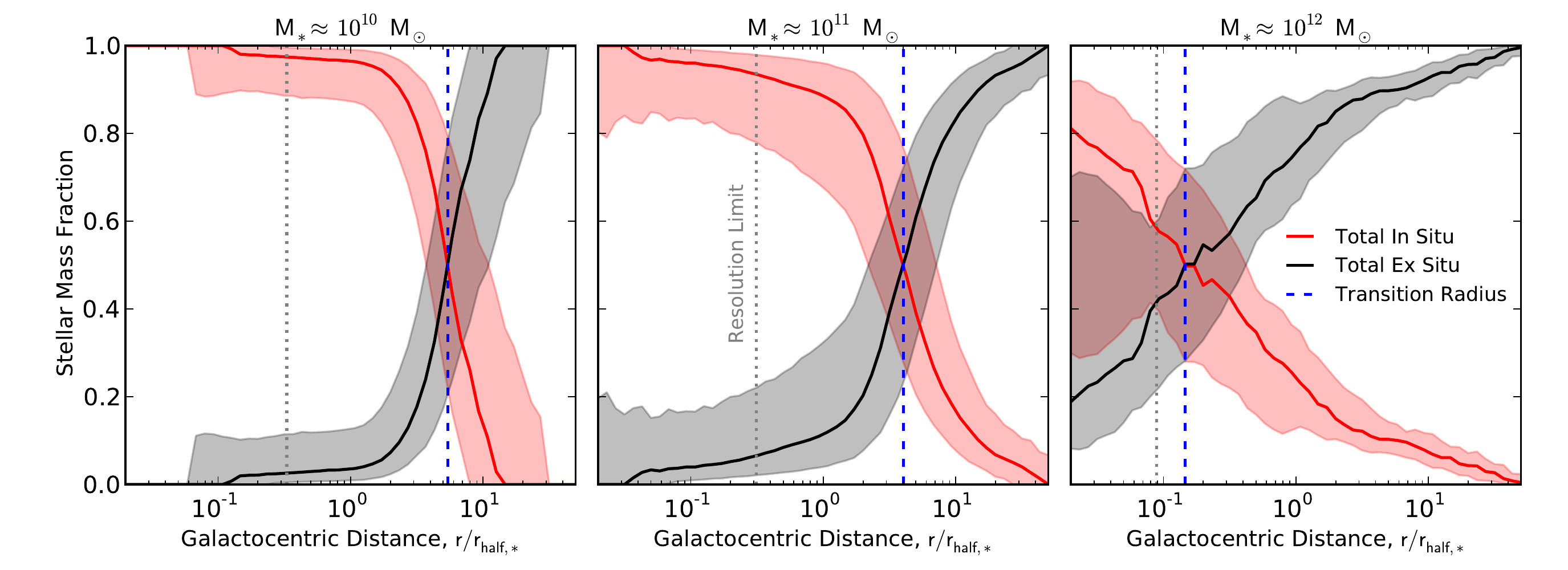}
		}
	}
\caption{Top: median stellar mass density profiles (averaged over spherical shells) of galaxies at $z=0$, shown as a function of galactocentric distance (in units of the stellar half-mass radius). The left, centre, and right panels correspond to the stellar mass bins centred at $10^{10}$, $10^{11}$ and $10^{12} \, \Msun$, respectively. The dot-dashed black line shows the total stellar content, while the solid lines correspond to stars that were formed in situ (red) and ex situ (black). Bottom: the median stellar mass fraction (averaged over spherical shells) as a function of normalized galactocentric distance, showing stars that were formed in situ (red) and ex situ (black). The intersection between the two lines, which we refer to as the \textit{normalized transition radius}, is indicated with a vertical dashed blue line. In all panels, the shaded regions indicate the 16th to 84th percentile range, or $1 \sigma$, while the dashed grey line shows the resolution limit (which corresponds to 4 softening lengths). This figure shows that the normalized transition radius changes with stellar mass, ranging from $\sim$4--5 effective radii for medium-sized galaxies to a fraction of an effective radius for very massive galaxies.}
\label{fig:f_acc_profiles}
\end{figure*}

\begin{figure*}
	\includegraphics[width=16cm]{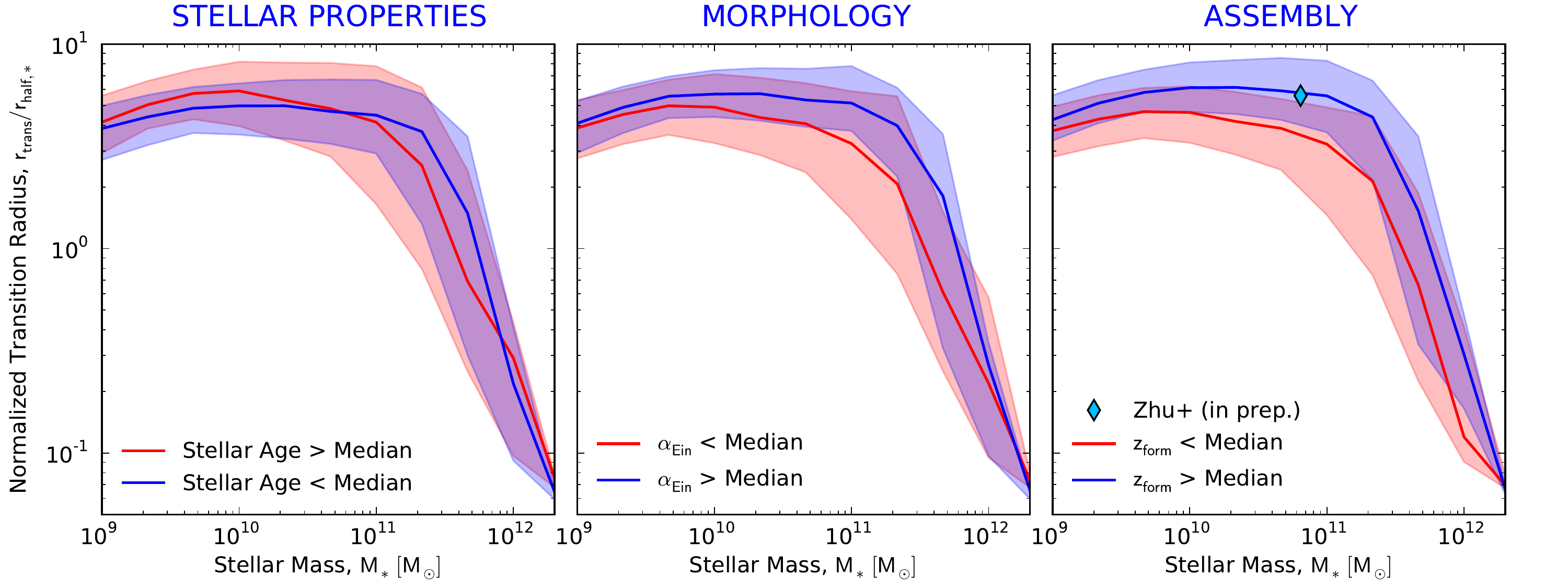}
	\caption{The median normalized transition radius as a function of stellar mass $M_{\ast}$, calculated for galaxies at $z=0$. The red and blue lines show the variation of the normalized transition radius with respect to different galaxy properties, at a fixed stellar mass. The dotted grey line shows the resolution limit, which is equal to 4 times the softening length, while the shaded regions indicate the 16th to 84th percentile ranges. A comparison between the current figure and the top panels from Fig. \ref{fig:f_acc_vs_mass_by_type} suggests that the normalized transition radius and the ex situ fraction are negatively correlated: at a fixed stellar mass, a higher (lower) ex situ fraction results in a smaller (larger) normalized transition radius.}
	\label{fig:r_cross_vs_mass_by_type}
\end{figure*}

\section{The spatial distribution of ex situ stars}\label{sec:spatial_distribution}

\begin{figure*}\centerline{
	\hbox{
	\includegraphics[width=7.5cm]{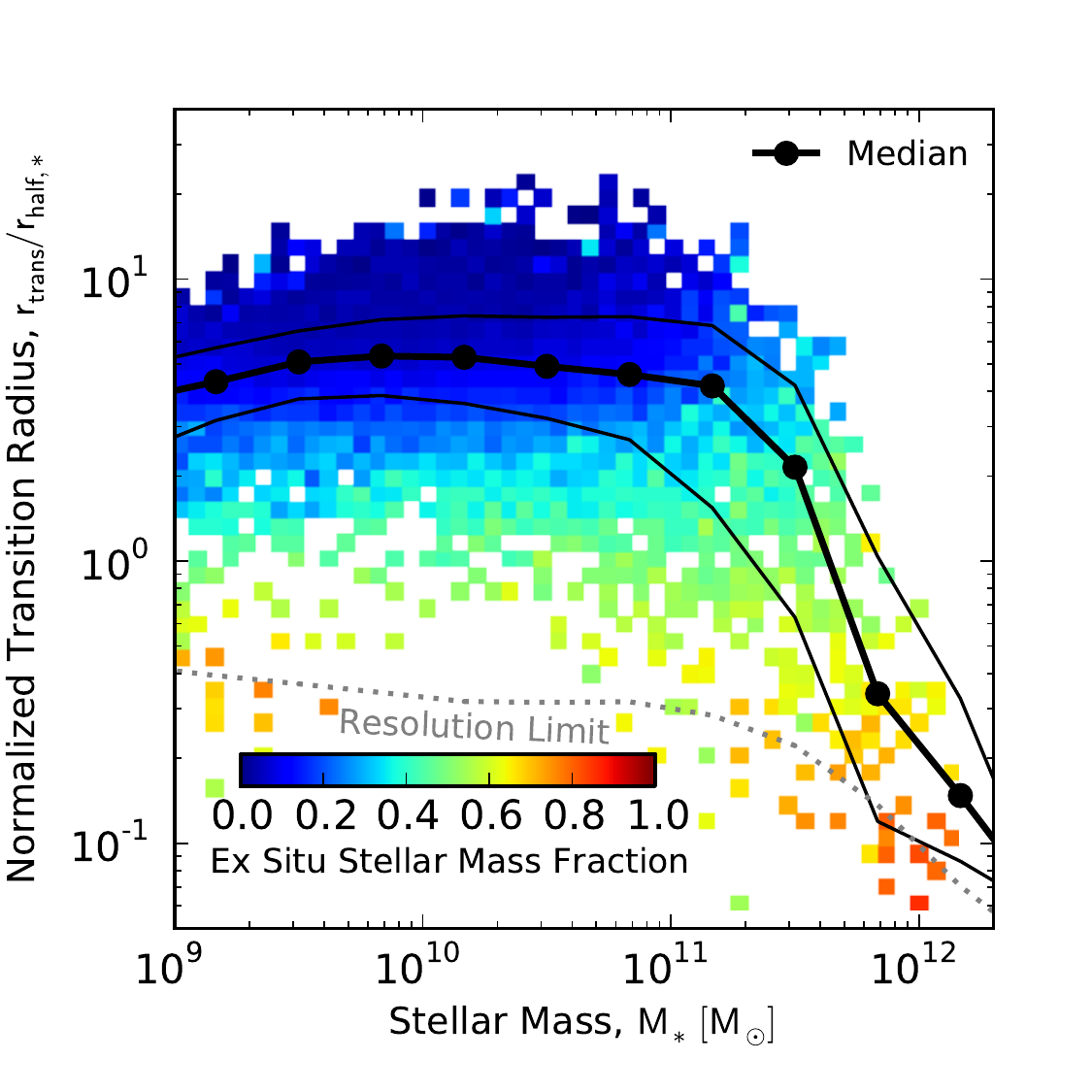}
	\includegraphics[width=7.5cm]{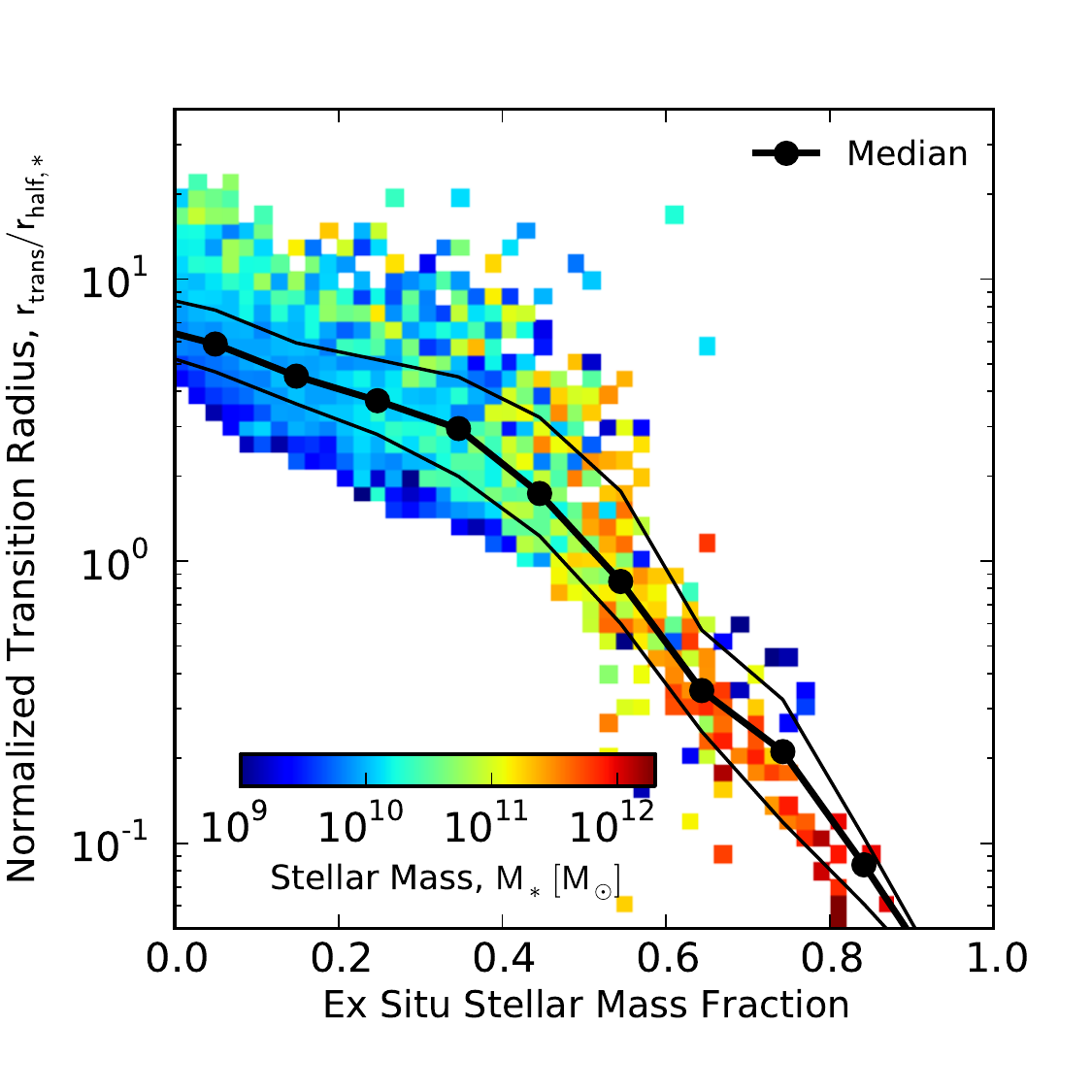}
	}}
	\caption{Left: The normalized transition radius as a function of stellar mass for galaxies at $z=0$. The colour of each bin corresponds to the median ex situ stellar mass fraction of the galaxies included in that bin. The dotted grey line shows the resolution limit, which is equal to 4 times the softening length. Right: The normalized transition radius as a function of the ex situ stellar mass fraction. Each bin is coloured according to the median stellar mass of the galaxies in the corresponding bin. In both panels, the thick black line shows the median trend, while the thin lines show the 16th and 84th percentiles. This figure shows that there is a strong, negative correlation between the normalized transition radius and the ex situ stellar mass fraction, and that the relationship between these two quantities is approximately independent of stellar mass.}
	\label{fig:transition_radius}
\end{figure*}

We have mentioned in Section \ref{subsec:general_trends} that the fraction of accreted stars measured within a sphere of twice the stellar half-mass radius is smaller than the corresponding fraction measured across the whole galaxy volume (as done throughout the paper), and that this trend becomes progressively more pronounced for less massive galaxies. This suggests a strong radial dependence in the distribution of in situ and ex situ stars, which we investigate in this section by focusing exclusively on $z=0$ galaxies.

\subsection{Visual impression and cumulative mass profiles}\label{subsec:visual_impression}

Fig. \ref{fig:projected_density} shows projected stellar density maps for a random selection of galaxies in three different mass bins (from top to bottom, total stellar masses of about $10^{10}$, $10^{11}$, and $10^{12} \, \Msun$, respectively), where the four different columns correspond to different stellar components. These images qualitatively demonstrate that the accreted material extends to much larger distances from the galactic centre than the in situ stars, producing the well known fine-structure features seen in galactic stellar haloes, such as streams and shells. On the other hand, in situ stars appear to reach larger densities towards the innermost regions of galaxies, i.e. close to their original birth sites. 

We quantify and expand on these statements in Fig. \ref{fig:f_acc_vs_radius_cumulative}, which shows the normalized, cumulative mass profiles $M_{\ast}(< r) / M_{\ast}$ of stars averaged across galaxies in different mass bins. The upper panels show the total in situ and ex situ components (red and black curves, respectively), while the lower panels show the different contributions of the accreted component, where blue, cyan, and green lines correspond to stars accreted in major ($\mu > 1/4$), minor ($1/10 < \mu < 1/4$), and very minor ($\mu < 1/10$) mergers, and the magenta line is used to represent stars that were stripped from surviving galaxies. By definition, the intersections with the horizontal dotted line (at a value of 0.5) correspond to the half-mass radii of the different components. We note that the ex situ half-mass radius is consistently found at 1.5--1.8 times the total stellar half-mass radius, and is larger than the in situ half-mass radius for galaxies of all masses, as expected from the accretion origin of the former \citep[e.g.][]{Searle1978, Abadi2006, Zolotov2009, Font2011, Pillepich2014, Pillepich2015}. 

Fig. \ref{fig:f_acc_vs_radius_cumulative} clearly demonstrates that in situ stars are found closest to the galactic centre, where most of the star-forming gas is located, followed by stars that were accreted from mergers in order of decreasing mass ratio (i.e. more minor mergers deposit their stellar content at larger radii), and finally stars that were captured from surviving galaxies (e.g. a flyby or an early passage of an ongoing merger), which were presumably very loosely bound to their former hosts. These trends are in agreement with numerical studies of galaxy mergers carried out in the last few decades. In particular, \cite{Barnes1988} and \cite{Hopkins2009} used various types of simulations to investigate non-dissipational and dissipational interactions between equal-mass galaxies (embedded in their respective DM haloes) and found that the binding energy ranking of particles is approximately conserved. As a corollary of this result, stars accreted from less massive galaxies -- which generally have smaller binding energies -- are expected to end up in the outskirts of the merger remnant. These findings are systematically reproduced and quantified in recent work by \cite{Amorisco2015}, who explores the deposition of stars in $N$-body merger remnants as a function of merger mass ratios, halo concentrations, and orbital properties.

\subsection{Differential profiles and the transition radius}

Fig. \ref{fig:f_acc_profiles} shows median radial profiles for the in situ and ex situ stellar components of galaxies with different masses at $z=0$, calculated over spherical shells. Again, the left, centre, and right panels correspond to galaxies with stellar masses of approximately $10^{10}$, $10^{11}$ and $10^{12} \, \Msun$, respectively. The top panels show the stellar mass density of the in situ (solid red), ex situ (solid black), and total (dot-dashed black) stellar components  as a function of galactocentric distance (normalized by the stellar half-mass radius of each galaxy). Three interesting observations can be noted by inspecting such panels. Firstly, the shapes of the in situ and ex situ profiles are different (more so at lower masses than in the highest mass bin): in particular, the in situ profiles appear to be dominated by disc structures in the innermost regions. Secondly, the galaxy-to-galaxy variations depicted by the shaded regions (corresponding to the range between the 16th and 84th percentiles) are different for the two components, being larger for the ex situ profiles than for the in situ ones (except for the $10^{12} \, \Msun$ galaxies). This is consistent with the scatter in the cumulative profiles from Fig. \ref{fig:f_acc_vs_radius_cumulative} and could be an indication that galaxies (especially those at the low-mass end, which undergo mergers less frequently) are more diverse in the way they merge and assemble their stellar mass than in the way they accrete gas and convert it into stars. Finally, there is always a particular radius at which the two components locally become equally abundant, which is smaller for more massive galaxies and which we refer to as the \textit{(normalized) transition radius}. This becomes even more clear in the lower panels, which show the fraction of in situ and ex situ stars (in spherical shells) as a function of galactocentric distance. The normalized transition radius is indicated with a vertical, dashed blue line. In all panels, we indicate the resolution limit (equal to 4 softening lengths) with a dotted grey line, so that the profiles should not necessarily be trusted below this scale.

Evidently, the normalized transition radius changes with stellar mass. We can also expect variations in the normalized transition radius with respect to other galaxy properties such as morphology and assembly history, but cannot predict a priori how large the differences will be. Therefore, we carry out a more systematic study of the transition radii as follows. For each galaxy, we compute its ex situ fraction as a function of galactocentric distance (in spherical shells), and then locate the transition radius by fitting a 5th order polynomial (with weights corresponding to the square root of the number of stellar particles in each spherical shell) and then finding the radius at which the ex situ fraction rises to a value of 0.5. We impose some restrictions, such as requiring that the cumulative number of particles found inside \textit{and} outside of the transition radius is at least 5. Sometimes the transition radius cannot be defined because a galaxy has an overwhelming amount of in situ or ex situ stellar particles at all radii, but these cases are extremely rare in the mass range considered ($M_{\ast} = 10^{9}-10^{12} \, \Msun$). The normalized transition radii are always given in units of the stellar half-mass radii of the corresponding galaxies.

Fig. \ref{fig:r_cross_vs_mass_by_type} shows the median normalized transition radius as a function of stellar mass for galaxies at $z=0$, in units of the stellar half mass radius. The three panels show the variation in the normalized transition radius with respect to different galaxy properties: stellar age (left), morphology (centre), and formation redshift (right). These quantities are the same as those in the top row from Fig. \ref{fig:f_acc_vs_mass_by_type} (the other six quantities yield similar results, so they are not shown in order to avoid repetition). The dotted grey line indicates the resolution limit (equal to 4 softening lengths) normalized by the median stellar half-mass radius of the galaxies in each mass bin. This figure shows that, at a fixed stellar mass, galaxies with more spheroidal morphologies and those associated with younger DM haloes exhibit slightly smaller transition radii, i.e. their stellar mass budget is dominated by ex situ stars down to smaller galactocentric distances.

An observational proxy for the transition radius has been investigated by \cite{DSouza2014}, who stacked a large number of galaxy images from Sloan Digital Sky Survey (SDSS) and studied them as a function of stellar mass and concentration. By fitting a multi-component S\'{e}rsic model to their stacked images, D'Souza et al. found that high-concentration galaxies have smaller transition radii than their low-concentration counterparts (at a fixed stellar mass), in qualitative agreement with this work.

A comparison between Fig. \ref{fig:r_cross_vs_mass_by_type} and the top panels of Fig. \ref{fig:f_acc_vs_mass_by_type} reveals a very clear trend: galaxy populations with higher (lower) ex situ fractions have smaller (larger) normalized transition radii. In order to explore this effect further, the left-hand panel from Fig. \ref{fig:transition_radius} shows the normalized transition radius as a function of stellar mass, with each two-dimensional bin coloured according to the median ex situ fraction of the galaxies included in that bin. The median and $1 \sigma$ scatter are indicated with thick and thin black lines, respectively. The predominantly horizontal distribution of the different colours suggests that there is a strong correlation between the normalized transition radius and the ex situ fraction, which is roughly independent of stellar mass. To verify this, the right panel from Fig. \ref{fig:transition_radius} shows the normalized transition radius as a function of the ex situ fraction, with bins coloured according to the median stellar mass. We observe a tight, negative correlation between the normalized transition radius and the ex situ fraction, as indicated by the thick and thin black lines, while the scatter shows a very weak correlation with stellar mass, in agreement with the horizontal colour arrangement from the left panel. This suggests that there is a nearly universal relationship between the normalized transition radius and the ex situ fraction.

The pattern just described \textit{could} be trivially explained if we could assume that the shapes of the in situ and ex situ density profiles (shown in the upper panels from Fig. \ref{fig:f_acc_profiles}) do not change with mass, such that an increase in ex situ stellar mass resulted in a vertical displacement of the ex situ density profile, bringing the normalized transition radius closer to the galactic centre in a systematic fashion. However, as we have already noted, the spherically-averaged profiles of both in situ and ex situ stars exhibit different shapes at different masses and galactocentric distances: hints of exponential discs can be seen in the in situ component within a few half-mass radii for small and medium-sized galaxies, while the slopes of the accreted material at large galactocentric distances become shallower for higher stellar masses (see \citealt{Pillepich2014} for a full description of the trends in the stellar halo density slopes).

\section{Discussion and conclusions}\label{sec:discussion_and_conclusions}

We have investigated several aspects of the stellar mass assembly of galaxies using data from the Illustris simulation, employing merger trees and a classification scheme for individual stellar particles as our main tools. In particular, we have investigated the contributions to the build-up of galaxies from the two channels of stellar mass assembly expected within the hierarchical growth of structures in a $\Lambda$CDM Universe: (1) in situ star formation, i.e. formation of stars occurring along the main progenitor branch of a given galaxy, and (2) ex situ mass growth, i.e. accretion of stars that formed within a galaxy other than the one under analysis and which were subsequently accreted. This is the first time, to our knowledge, that such an analysis has been carried out directly on a large-scale, hydrodynamic cosmological simulation over a wide range of galaxy masses ($M_{\ast}  = 10^9-10^{12} \Msun$), galaxy types, environments, and DM halo assembly histories.

We begin by quantifying the \textit{specific stellar mass accretion rate} of galaxies, which measures the average amount of stellar mass accreted through mergers \textit{per unit time} as a function of descendant stellar mass $M_0$, merger mass ratio $\mu$, and redshift $z$. This quantity is very closely related to the galaxy-galaxy merger rate, which has been determined to great accuracy in \cite{Rodriguez-Gomez2015}, and is also very similar to the dimensionless galaxy growth rate (from mergers) previously studied by \cite{Guo2008}. On average, we find that the contributions to stellar mass growth due to different types of mergers are 50--60 per cent from major mergers ($\mu > 1/4$), 20--25 per cent from minor mergers ($1/10 < \mu < 1/4$), and 20--25 per cent from very minor mergers ($\mu < 1/10$). We provide a fitting formula for the specific stellar mass accretion rate which is accurate over a wide range of stellar masses, merger mass ratios, and redshifts. The mathematical form of this fitting function reveals that the specific stellar mass accretion rate has a much stronger mass dependence than the specific DM accretion rate of DM haloes obtained from $N$-body simulations \citep[e.g.][]{Fakhouri2010, Genel2010}.

Moreover, the specific stellar mass accretion rate can be directly compared to the specific star formation rate. This comparison shows that the stellar mass growth of most galaxies is dominated by in situ star formation, except for sufficiently massive galaxies at $z \lesssim 1$, which grow primarily by mergers due to their higher specific stellar mass accretion rates and lower specific star formation rates. In particular, at $z \approx 0.1$ a transition point between the two modes of stellar mass growth occurs for galaxies with $M_{\ast} \approx$ 1--2 $\times 10^{11} \, \Msun$ (Fig. \ref{fig:macc_vs_mass}).

Besides quantifying the instantaneous rate of growth due to mergers, we also investigated the `time-integrated' amount of accreted stars and their spatial distribution. The \textit{ex situ stellar mass fraction} measures the fraction of the total stellar mass in a galaxy contributed by stars that formed in other galaxies and which were accreted later on. In order to determine this quantity accurately, we use the merger trees to investigate the origin of individual stellar particles in the simulation, classifying them as \textit{in situ} or \textit{ex situ}, along with other useful information such as the mass ratio of the mergers in which the particles were accreted (when applicable). We find that some ex situ stars cannot be associated with a (completed) merger event, but that instead they were \textit{stripped from surviving galaxies}, as can happen during a flyby event or during close passages of an ongoing merger.

By using the stellar particle classification scheme described above, we find that more massive galaxies have a larger fraction of accreted stars, with values at $z=0$ ranging from $\sim$10 per cent for a typical Milky Way-sized galaxy (without taking its merging history into account) to over 80 per cent for the most massive galaxies in the simulation ($M_{\ast} \approx 10^{12} \Msun$), yet with large scatter at a fixed stellar mass. Compared to previous theoretical works \citep[e.g.][]{Oser2010, Lackner2012a, Lee2013, Pillepich2015, Hirschmann2015}, our determination appears to lie close to the `median' prediction from these other studies, reducing the tension between them (Fig. \ref{fig:ex_situ_main}). Importantly, we show that the ex situ stellar mass fraction at $z=0$ shows excellent convergence when obtained from simulations with different resolutions (Fig. \ref{fig:ex_situ_main}) or when calculated with different methodologies (Fig. \ref{fig:ex_situ_allinone}, top-left).

Since our stellar particle classification scheme also contains information about the mass ratios of the mergers in which individual ex situ stellar particles were accreted, we can decompose the ex situ stellar mass of a galaxy according to the origin of its different `components:' stars accreted in major, minor, and very minor mergers, as well as stars that were stripped from surviving galaxies (such as flybys or mergers in progress). The latter component cannot be accounted for when using merger trees alone. On average, major mergers are responsible for $\sim$50 per cent of the ex situ stellar mass in galaxies at $z=0$, while minor and very minor mergers contribute $\sim$20 per cent each, and finally stars that were stripped from surviving galaxies contribute another 10 per cent (Fig. \ref{fig:ex_situ_allinone}, bottom-right). To a first approximation, these fractions are independent of stellar mass, in agreement with our findings for the galaxy-galaxy merger rate \citep{Rodriguez-Gomez2015} and the specific stellar mass accretion rate (Section \ref{sec:stellar_mass_accretion}).

At a fixed stellar mass, the contribution of the accreted stars to the total stellar mass exhibits a non-negligible scatter from galaxy to galaxy, with $1 \sigma$ variations as large as factors of a few for galaxies in the $10^{10}$--$10^{11} \Msun$ mass range. We investigate the origin of such scatter by separating galaxies according to different galactic and halo properties (at a fixed stellar mass) and then comparing the ex situ stellar mass fractions of the two resulting galaxy populations. We find that the residual dependence on stellar properties such as stellar age, colour, and star formation rate is weak, whereas morphology, halo formation time, and recent merging history have a definite impact on the ex situ fraction -- namely, galaxies of the same stellar mass that have spheroidal morphologies, late halo formation times, or violent recent merging histories also tend to have higher ex situ fractions (Fig. \ref{fig:f_acc_vs_mass_by_type}). 

Beyond the correlation with stellar mass and various galactic and halo properties at $z=0$, we also examine the redshift evolution of the ex situ stellar mass fraction. We begin by considering all Illustris galaxies as a whole (Fig. \ref{fig:f_acc_vs_z_snaps}, left), which reveals that the \textit{global} ex situ stellar mass fraction increases monotonically with time, reaching 17 per cent at $z=2$ and 30 per cent at $z=0$. This means that the majority of stars in the Universe at any redshift are located close to their original birth sites. When focusing on individual systems, we find that the stellar content of most galaxies at any redshift was predominantly formed in situ, except for the most massive galaxies in our simulation ($\sim 10^{12} \Msun$ at $z=0$), for which we find a `crossover' between the in situ and ex situ stellar components at $z \approx 1$ (Fig. \ref{fig:f_acc_vs_z}). This is consistent with our findings from Section \ref{sec:stellar_mass_accretion} based on instantaneous growth rates and means that the so-called `two-phase' model of galaxy formation is only a good approximation for our most massive galaxies, while \cite{Oser2010} found a `crossover' for all the galaxies they analysed ($ M_{\ast} \approx 4$--$40 \times 10^{10} \, h^{-1} \Msun$). The differences with respect to Oser et al. can be partially ascribed to different definitions (e.g. their in situ stars were simply defined as those formed within $0.1 R_{\rm vir}$), but more importantly to their different feedback implementation, which resulted in a substantial amount of early star formation, possibly overestimating the amount of stellar accretion at later times for all their galaxies. Our results are in better qualitative agreement with predictions from semi-empirical models \citep{Moster2012a}, which also determine that in situ stars should be the dominant stellar component in galaxies over a wide range of masses, as well as with hydrodynamic simulations of individual late-type galaxies \citep{Pillepich2015}.

Having characterized the overall growth of the stellar mass of a galaxy due to in situ star formation and stellar accretion, both in terms of instantaneous rates as well as `time-integrated' mass fractions, we proceed to investigate the spatial distribution of the stellar components that result from such growth channels. A visual assessment of the spatial distribution of in situ and ex situ stars shows that the former are typically found closer to the galactic centre, while ex situ stars have a more extended spatial distribution, are less concentrated toward the galactic centre, and can exhibit stellar halo features such as streams and shells (Fig. \ref{fig:projected_density}).

In general, the stellar content of a galaxy displays the following spatial segregation: in situ stars are found closest to the galactic centre, followed by stars accreted in major, minor, and very minor mergers (in the same order), and finally stars that were stripped from surviving galaxies, which are found at the largest galactocentric distances (Fig. \ref{fig:f_acc_vs_radius_cumulative}). This is consistent with expectations from $N$-body and hydrodynamic simulations of galaxy mergers, in which the binding energy ranks of the interacting particles tend to be conserved during a galaxy merger \citep[e.g.][]{Amorisco2015, Barnes1988, Hopkins2009}.

In fact, it is well known that in situ stars tend to be concentrated toward the centre of the galaxy, close to their original formation sites, while the spatial distribution of ex situ stars is generally more extended, the latter being the dominant component of galactic stellar haloes \citep[e.g.][]{Searle1978, Abadi2006, Zolotov2009, Font2011, Pillepich2014, Pillepich2015}. This implies the existence of a `transition' radius, defined as the distance from the galactic centre where in situ and ex situ stars locally become equally abundant (Fig. \ref{fig:f_acc_profiles}). We find that the \textit{normalized transition radius} (given in units of the stellar half-mass radius, $r_{\rm half}$) shows the following trends at $z=0$ with respect to stellar mass: it is relatively flat at 4--5 $r_{\rm half}$ for medium-sized galaxies ($M_{\ast} \approx 10^{10}-10^{11} \, \Msun$) after which it declines sharply, reaching $\sim$1 $r_{\rm half}$ at $M_{\ast} \approx 5 \times 10^{11} \Msun$ and $\sim$0.2 $r_{\rm half}$ at $M_{\ast} \approx 10^{12} \Msun$. We also quantify the correlation between the normalized transition radius with galactic and halo properties. At a fixed stellar mass, the normalized transition radius is smaller for galaxies that have spheroidal morphologies, late halo formation times, and violent recent merging histories (Fig. \ref{fig:r_cross_vs_mass_by_type}). A comparison of these trends with Fig. \ref{fig:f_acc_vs_mass_by_type} suggests that the normalized transition radius is negatively correlated with the ex situ stellar mass fraction, i.e. that galaxies with higher (lower) ex situ fractions also tend to have smaller (larger) transition radii, as we explicitly demonstrate in Fig. \ref{fig:transition_radius}. Although such a trend between the ex situ stellar mass fraction and the normalized transition radius is not unexpected, it is interesting that the two quantities follow a nearly universal relation that is approximately independent of stellar mass.

Theoretical predictions of the transition radius can be useful in guiding observations of stellar haloes. For example, it is believed that the oldest and most metal-poor stars of a Milky Way-sized galaxy are the ex situ stars found in its stellar halo, which would therefore contain clues about the formation of the galaxy and its former interactions with other objects. However, observations are unable to distinguish in situ stars from ex situ ones, since the chemical and kinematic properties of the two stellar populations do exhibit significant overlap \citep[e.g.][]{Pillepich2015}, i.e. observations cannot truly assess the amount of `contamination' from in situ stars. For this reason, knowledge of the transition radius (or, more generally, predictions for the overall spatial distribution of accreted and in situ stars) can help in targeting regions of stellar haloes which are largely devoid of in situ stars, but which at the same time are not too faint to observe, i.e. a `sweet spot' in galactocentric distance. We note, however, that the transition radii presented in this work were obtained by taking spherical averages of the stellar particle distributions. Therefore, in the case of a disc-like galaxy, the transition radius should be interpreted as a measure of the extent of the galactic disc, and the transition point at which the local densities of in situ and ex situ stars become equivalent should happen at a smaller galactocentric distance if measured at high latitudes.

The angular distribution of accreted stars, which is crucial for informing observations and theoretical models of the Milky Way stellar halo, will be addressed in further work, along with an investigation of the metallicity and stellar age gradients of Illustris galaxies in the context of in situ and ex situ stellar populations. In general, we expect our results to be very helpful in understanding diverse aspects of the formation of galaxies and their stellar haloes, in combination with data from upcoming deep and wide-field galaxy surveys.

\section*{Acknowledgements}

We thank Jeremiah P. Ostriker, Gurtina Besla, and Benedikt Diemer for useful comments and discussions. SG acknowledges support provided by NASA through Hubble Fellowship grant HST-HF2-51341.001-A awarded by the STScI, which is operated by the Association of Universities for Research in Astronomy, Inc., for NASA, under contract NAS5-26555. VS acknowledges support through the European Research Council through ERC-StG grant EXAGAL-308037. LH acknowledges support from NSF grant AST-1312095 and NASA grant NNX12AC67G. Simulations were run on the Harvard Odyssey and CfA/ITC clusters, the Ranger and Stampede supercomputers at the Texas Advanced Computing Center as part of XSEDE, the Kraken supercomputer at Oak Ridge National Laboratory as part of XSEDE, the CURIE supercomputer at CEA/France as part of PRACE project RA0844, and the SuperMUC computer at the Leibniz Computing Centre, Germany, as part of project pr85je.

\bibliographystyle{mn2eFixed}

\bibliography{paper}

\begin{thebibliography}{88}
\expandafter\ifx\csname natexlab\endcsname\relax\def\natexlab#1{#1}\fi

\bibitem[{Abadi, Navarro \& Steinmetz(2006)Abadi, Navarro, \&
  Steinmetz}]{Abadi2006}
Abadi M.~G., Navarro J.~F., Steinmetz M., 2006, MNRAS, 365, 747

\bibitem[{Abadi {et~al}\mbox{.}(2003)Abadi, Navarro, Steinmetz, \&
  Eke}]{Abadi2003}
Abadi M.~G., Navarro J.~F., Steinmetz M., Eke V.~R., 2003, ApJ, 597, 21

\bibitem[{Amorisco(2015)}]{Amorisco2015}
Amorisco N.~C., 2015, preprint (arXiv:1511.08806)

\bibitem[{Barnes(1988)}]{Barnes1988}
Barnes J.~E., 1988, ApJ, 331, 699

\bibitem[{Behroozi {et~al}\mbox{.}(2013)Behroozi, Marchesini, Wechsler, Muzzin,
  Papovich, \& Stefanon}]{Behroozi2013}
Behroozi P.~S., Marchesini D., Wechsler R.~H., Muzzin A., Papovich C., Stefanon
  M., 2013, ApJ, 777, L10

\bibitem[{Behroozi, Wechsler \& Conroy(2013)Behroozi, Wechsler, \&
  Conroy}]{Behroozi2013b}
Behroozi P.~S., Wechsler R.~H., Conroy C., 2013, ApJ, 770, 57

\bibitem[{Bird {et~al}\mbox{.}(2014)Bird, Vogelsberger, Haehnelt, Sijacki,
  Genel, Torrey, Springel, \& Hernquist}]{Bird2014}
Bird S., Vogelsberger M., Haehnelt M., Sijacki D., Genel S., Torrey P.,
  Springel V., Hernquist L., 2014, MNRAS, 445, 2313

\bibitem[{Bray {et~al}\mbox{.}(2016)Bray, Pillepich, Sales, Zhu, Genel,
  Rodriguez-Gomez, Torrey, Nelson, Vogelsberger, Springel, Eisenstein, \&
  Hernquist}]{Bray2016}
Bray A.~D. {et~al.}, 2016, MNRAS, 455, 185

\bibitem[{Bundy, Treu \& Ellis(2007)Bundy, Treu, \& Ellis}]{Bundy2007}
Bundy K., Treu T., Ellis R.~S., 2007, ApJ, 665, L5

\bibitem[{Cimatti, Nipoti \& Cassata(2012)Cimatti, Nipoti, \&
  Cassata}]{Cimatti2012}
Cimatti A., Nipoti C., Cassata P., 2012, MNRAS, 422, L62

\bibitem[{Cooper {et~al}\mbox{.}(2010)Cooper, Cole, Frenk, White, Helly,
  Benson, {De Lucia}, Helmi, Jenkins, Navarro, Springel, \& Wang}]{Cooper2010}
Cooper A.~P. {et~al.}, 2010, MNRAS, 406, 744

\bibitem[{Cooper {et~al}\mbox{.}(2013)Cooper, D'Souza, Kauffmann, Wang,
  Boylan-Kolchin, Guo, Frenk, \& White}]{Cooper2013}
Cooper A.~P., D'Souza R., Kauffmann G., Wang J., Boylan-Kolchin M., Guo Q.,
  Frenk C.~S., White S. D.~M., 2013, MNRAS, 434, 3348

\bibitem[{Cooper {et~al}\mbox{.}(2015)Cooper, Parry, Lowing, Cole, \&
  Frenk}]{Cooper2015}
Cooper A.~P., Parry O.~H., Lowing B., Cole S., Frenk C., 2015, MNRAS, 454, 3185

\bibitem[{Daddi {et~al}\mbox{.}(2005)Daddi, Renzini, Pirzkal, Cimatti,
  Malhotra, Stiavelli, Xu, Pasquali, Rhoads, Brusa, {di Serego Alighieri},
  Ferguson, Koekemoer, Moustakas, Panagia, \& Windhorst}]{Daddi2005}
Daddi E. {et~al.}, 2005, ApJ, 626, 680

\bibitem[{Damjanov {et~al}\mbox{.}(2009)Damjanov, McCarthy, Abraham,
  Glazebrook, Yan, Mentuch, {Le Borgne}, Savaglio, Crampton, Murowinski,
  Juneau, Carlberg, J{\o}rgensen, Roth, Chen, \& Marzke}]{Damjanov2009}
Damjanov I. {et~al.}, 2009, ApJ, 695, 101

\bibitem[{Davis {et~al}\mbox{.}(1985)Davis, Efstathiou, Frenk, \&
  White}]{Davis1985}
Davis M., Efstathiou G., Frenk C.~S., White S. D.~M., 1985, ApJ, 292, 371

\bibitem[{{De Lucia} \& Blaizot(2007)}]{DeLucia2007}
{De Lucia} G., Blaizot J., 2007, MNRAS, 375, 2

\bibitem[{{De Lucia} {et~al}\mbox{.}(2011){De Lucia}, Fontanot, Wilman, \&
  Monaco}]{DeLucia2011}
{De Lucia} G., Fontanot F., Wilman D., Monaco P., 2011, MNRAS, 414, 1439

\bibitem[{Deason, Belokurov \& Evans(2011)Deason, Belokurov, \&
  Evans}]{Deason2011}
Deason A.~J., Belokurov V., Evans N.~W., 2011, MNRAS, 416, 2903

\bibitem[{Deason {et~al}\mbox{.}(2013)Deason, Belokurov, Evans, \&
  Johnston}]{Deason2013}
Deason A.~J., Belokurov V., Evans N.~W., Johnston K.~V., 2013, ApJ, 763, 113

\bibitem[{Dekel {et~al}\mbox{.}(2009)Dekel, Birnboim, Engel, Freundlich,
  Goerdt, Mumcuoglu, Neistein, Pichon, Teyssier, \& Zinger}]{Dekel2009}
Dekel A. {et~al.}, 2009, Nature, 457, 451

\bibitem[{Dolag {et~al}\mbox{.}(2009)Dolag, Borgani, Murante, \&
  Springel}]{Dolag2009a}
Dolag K., Borgani S., Murante G., Springel V., 2009, MNRAS, 399, 497

\bibitem[{D'Souza {et~al}\mbox{.}(2014)D'Souza, Kauffman, Wang, \&
  Vegetti}]{DSouza2014}
D'Souza R., Kauffman G., Wang J., Vegetti S., 2014, MNRAS, 443, 1433

\bibitem[{Dubois {et~al}\mbox{.}(2013)Dubois, Gavazzi, Peirani, \&
  Silk}]{Dubois2013}
Dubois Y., Gavazzi R., Peirani S., Silk J., 2013, MNRAS, 433, 3297

\bibitem[{Eggen, Lynden-Bell \& Sandage(1962)Eggen, Lynden-Bell, \&
  Sandage}]{Eggen1962}
Eggen O.~J., Lynden-Bell D., Sandage A.~R., 1962, ApJ, 136, 748

\bibitem[{Fakhouri \& Ma(2008)}]{Fakhouri2008}
Fakhouri O., Ma C.-P., 2008, MNRAS, 386, 577

\bibitem[{Fakhouri, Ma \& Boylan-Kolchin(2010)Fakhouri, Ma, \&
  Boylan-Kolchin}]{Fakhouri2010}
Fakhouri O., Ma C.-P., Boylan-Kolchin M., 2010, MNRAS, 406, 2267

\bibitem[{Feldmann {et~al}\mbox{.}(2010)Feldmann, Carollo, Mayer, Renzini,
  Lake, Quinn, Stinson, \& Yepes}]{Feldmann2010}
Feldmann R., Carollo C.~M., Mayer L., Renzini A., Lake G., Quinn T., Stinson
  G.~S., Yepes G., 2010, ApJ, 709, 218

\bibitem[{Fiacconi, Feldmann \& Mayer(2014)Fiacconi, Feldmann, \&
  Mayer}]{Fiacconi2014}
Fiacconi D., Feldmann R., Mayer L., 2014, MNRAS, 446, 1957

\bibitem[{Font {et~al}\mbox{.}(2011)Font, McCarthy, Crain, Theuns, Schaye,
  Wiersma, \& Vecchia}]{Font2011}
Font A.~S., McCarthy I.~G., Crain R.~A., Theuns T., Schaye J., Wiersma R.
  P.~C., Vecchia C.~D., 2011, MNRAS, 416, 2802

\bibitem[{Foreman-Mackey {et~al}\mbox{.}(2013)Foreman-Mackey, Hogg, Lang, \&
  Goodman}]{Foreman-Mackey2013}
Foreman-Mackey D., Hogg D.~W., Lang D., Goodman J., 2013, PASP, 125, 306

\bibitem[{Genel {et~al}\mbox{.}(2010)Genel, Bouch{\'{e}}, Naab, Sternberg, \&
  Genzel}]{Genel2010}
Genel S., Bouch{\'{e}} N., Naab T., Sternberg A., Genzel R., 2010, ApJ, 719,
  229

\bibitem[{Genel {et~al}\mbox{.}(2015)Genel, Fall, Hernquist, Vogelsberger,
  Snyder, Rodriguez-Gomez, Sijacki, \& Springel}]{Genel2015}
Genel S., Fall S.~M., Hernquist L., Vogelsberger M., Snyder G.~F.,
  Rodriguez-Gomez V., Sijacki D., Springel V., 2015, ApJ, 804, L40

\bibitem[{Genel {et~al}\mbox{.}(2009)Genel, Genzel, Bouch{\'{e}}, Naab, \&
  Sternberg}]{Genel2009}
Genel S., Genzel R., Bouch{\'{e}} N., Naab T., Sternberg A., 2009, ApJ, 701,
  2002

\bibitem[{Genel {et~al}\mbox{.}(2014)Genel, Vogelsberger, Springel, Sijacki,
  Nelson, Snyder, Rodriguez-Gomez, Torrey, \& Hernquist}]{Genel2014a}
Genel S. {et~al.}, 2014, MNRAS, 445, 175

\bibitem[{Greene {et~al}\mbox{.}(2015)Greene, Janish, Ma, McConnell, Blakeslee,
  Thomas, \& Murphy}]{Greene2015}
Greene J.~E., Janish R., Ma C.-P., McConnell N.~J., Blakeslee J.~P., Thomas J.,
  Murphy J.~D., 2015, ApJ, 807, 11

\bibitem[{Guo \& White(2008)}]{Guo2008}
Guo Q., White S. D.~M., 2008, MNRAS, 384, 2

\bibitem[{Hinshaw {et~al}\mbox{.}(2013)Hinshaw, Larson, Komatsu, Spergel,
  Bennett, Dunkley, Nolta, Halpern, Hill, Odegard, Page, Smith, Weiland, Gold,
  Jarosik, Kogut, Limon, Meyer, Tucker, Wollack, \& Wright}]{Hinshaw2013}
Hinshaw G. {et~al.}, 2013, ApJS, 208, 19

\bibitem[{Hirschmann {et~al}\mbox{.}(2015)Hirschmann, Naab, Ostriker, Forbes,
  Duc, Dave, Oser, \& Karabal}]{Hirschmann2015}
Hirschmann M., Naab T., Ostriker J.~P., Forbes D.~A., Duc P.-A., Dave R., Oser
  L., Karabal E., 2015, MNRAS, 449, 528

\bibitem[{Hopkins {et~al}\mbox{.}(2010{\natexlab{a}})Hopkins, Bundy, Croton,
  Hernquist, Keres, Khochfar, Stewart, Wetzel, \& Younger}]{Hopkins2010}
Hopkins P.~F. {et~al.}, 2010{\natexlab{a}}, ApJ, 715, 202

\bibitem[{Hopkins {et~al}\mbox{.}(2010{\natexlab{b}})Hopkins, Bundy, Hernquist,
  Wuyts, \& Cox}]{Hopkins2010b}
Hopkins P.~F., Bundy K., Hernquist L., Wuyts S., Cox T.~J., 2010{\natexlab{b}},
  MNRAS, 401, 1099

\bibitem[{Hopkins {et~al}\mbox{.}(2009)Hopkins, Lauer, Cox, Hernquist, \&
  Kormendy}]{Hopkins2009}
Hopkins P.~F., Lauer T.~R., Cox T.~J., Hernquist L., Kormendy J., 2009, ApJS,
  181, 486

\bibitem[{Hopkins {et~al}\mbox{.}(2010{\natexlab{c}})Hopkins, Younger, Hayward,
  Narayanan, \& Hernquist}]{Hopkins2010a}
Hopkins P.~F., Younger J.~D., Hayward C.~C., Narayanan D., Hernquist L.,
  2010{\natexlab{c}}, MNRAS, 402, 1693

\bibitem[{Jim{\'{e}}nez {et~al}\mbox{.}(2011)Jim{\'{e}}nez, Cora, Bassino,
  Tecce, \& {Smith Castelli}}]{Jimenez2011}
Jim{\'{e}}nez N., Cora S.~A., Bassino L.~P., Tecce T.~E., {Smith Castelli}
  A.~V., 2011, MNRAS, 417, 785

\bibitem[{Kere{\v{s}} {et~al}\mbox{.}(2005)Kere{\v{s}}, Katz, Weinberg, \&
  Dav{\'{e}}}]{Keres2005}
Kere{\v{s}} D., Katz N., Weinberg D.~H., Dav{\'{e}} R., 2005, MNRAS, 363, 2

\bibitem[{Lackner {et~al}\mbox{.}(2012)Lackner, Cen, Ostriker, \&
  Joung}]{Lackner2012a}
Lackner C.~N., Cen R., Ostriker J.~P., Joung M.~R., 2012, MNRAS, 425, 641

\bibitem[{Lee \& Yi(2013)}]{Lee2013}
Lee J., Yi S.~K., 2013, ApJ, 766, 38

\bibitem[{L{\'{o}}pez-Sanjuan {et~al}\mbox{.}(2012)L{\'{o}}pez-Sanjuan, {Le
  F{\`{e}}vre}, Ilbert, Tasca, Bridge, Cucciati, Kampczyk, Pozzetti, Xu,
  Carollo, Contini, Kneib, Lilly, Mainieri, Renzini, Sanders, Scodeggio,
  Scoville, Taniguchi, Zamorani, Aussel, Bardelli, Bolzonella, Bongiorno,
  Capak, Caputi, de~la Torre, de~Ravel, Franzetti, Garilli, Iovino, Knobel,
  Kova{\v{c}}, Lamareille, {Le Borgne}, {Le Brun}, {Le Floc’h}, Maier,
  McCracken, Mignoli, Pell{\'{o}}, Peng, P{\'{e}}rez-Montero, Presotto,
  Ricciardelli, Salvato, Silverman, Tanaka, Tresse, Vergani, Zucca, Barnes,
  Bordoloi, Cappi, Cimatti, Coppa, Koekemoer, Liu, Moresco, Nair, Oesch,
  Schawinski, \& Welikala}]{Lopez-Sanjuan2012}
L{\'{o}}pez-Sanjuan C. {et~al.}, 2012, A{\&}A, 548, A7

\bibitem[{Marinacci, Pakmor \& Springel(2014)Marinacci, Pakmor, \&
  Springel}]{Marinacci2013}
Marinacci F., Pakmor R., Springel V., 2014, MNRAS, 437, 1750

\bibitem[{McCarthy {et~al}\mbox{.}(2012)McCarthy, Font, Crain, Deason, Schaye,
  \& Theuns}]{McCarthy2012}
McCarthy I.~G., Font A.~S., Crain R.~A., Deason A.~J., Schaye J., Theuns T.,
  2012, MNRAS, 420, 2245

\bibitem[{McMillan(2011)}]{McMillan2011}
McMillan P.~J., 2011, MNRAS, 414, 2446

\bibitem[{Mistani {et~al}\mbox{.}(2016)Mistani, Sales, Pillepich,
  Sanchez-Janssen, Vogelsberger, Nelson, Rodriguez-Gomez, Torrey, \&
  Hernquist}]{Mistani2016}
Mistani P.~a. {et~al.}, 2016, MNRAS, 455, 2323

\bibitem[{Moster, Naab \& White(2013)Moster, Naab, \& White}]{Moster2012a}
Moster B.~P., Naab T., White S. D.~M., 2013, MNRAS, 428, 3121

\bibitem[{Naab, Johansson \& Ostriker(2009)Naab, Johansson, \&
  Ostriker}]{Naab2009}
Naab T., Johansson P.~H., Ostriker J.~P., 2009, ApJ, 699, L178

\bibitem[{Nelson {et~al}\mbox{.}(2015)Nelson, Pillepich, Genel, Vogelsberger,
  Springel, Torrey, Rodriguez-Gomez, Sijacki, Snyder, Griffen, Marinacci,
  Blecha, Sales, Xu, \& Hernquist}]{Nelson2015}
Nelson D. {et~al.}, 2015, A{\&}C, 13, 12

\bibitem[{Nelson {et~al}\mbox{.}(2013)Nelson, Vogelsberger, Genel, Sijacki,
  Kere{\v{s}}, Springel, \& Hernquist}]{Nelson2013}
Nelson D., Vogelsberger M., Genel S., Sijacki D., Kere{\v{s}} D., Springel V.,
  Hernquist L., 2013, MNRAS, 429, 3353

\bibitem[{Oesch {et~al}\mbox{.}(2010)Oesch, Carollo, Feldmann, Hahn, Lilly,
  Sargent, Scarlata, Aller, Aussel, Bolzonella, Bschorr, Bundy, Capak, Ilbert,
  Kneib, Koekemoer, Kova{\v{c}}, Leauthaud, {Le Floc'h}, Massey, McCracken,
  Pozzetti, Renzini, Rhodes, Salvato, Sanders, Scoville, Sheth, Taniguchi, \&
  Thompson}]{Oesch2010}
Oesch P.~A. {et~al.}, 2010, ApJ, 714, L47

\bibitem[{Oser {et~al}\mbox{.}(2012)Oser, Naab, Ostriker, \&
  Johansson}]{Oser2012}
Oser L., Naab T., Ostriker J.~P., Johansson P.~H., 2012, ApJ, 744, 63

\bibitem[{Oser {et~al}\mbox{.}(2010)Oser, Ostriker, Naab, Johansson, \&
  Burkert}]{Oser2010}
Oser L., Ostriker J.~P., Naab T., Johansson P.~H., Burkert A., 2010, ApJ, 725,
  2312

\bibitem[{Pastorello {et~al}\mbox{.}(2014)Pastorello, Forbes, Foster, Brodie,
  Usher, Romanowsky, Strader, \& Arnold}]{Pastorello2014}
Pastorello N., Forbes D.~a., Foster C., Brodie J.~P., Usher C., Romanowsky
  A.~J., Strader J., Arnold J.~a., 2014, MNRAS, 442, 1003

\bibitem[{Pillepich, Madau \& Mayer(2015)Pillepich, Madau, \&
  Mayer}]{Pillepich2015}
Pillepich A., Madau P., Mayer L., 2015, ApJ, 799, 184

\bibitem[{Pillepich {et~al}\mbox{.}(2014)Pillepich, Vogelsberger, Deason,
  Rodriguez-Gomez, Genel, Nelson, Torrey, Sales, Marinacci, Springel, Sijacki,
  \& Hernquist}]{Pillepich2014}
Pillepich A. {et~al.}, 2014, MNRAS, 444, 237

\bibitem[{Rashkov {et~al}\mbox{.}(2012)Rashkov, Madau, Kuhlen, \&
  Diemand}]{Rashkov2012}
Rashkov V., Madau P., Kuhlen M., Diemand J., 2012, ApJ, 745, 142

\bibitem[{Robotham {et~al}\mbox{.}(2014)Robotham, Driver, Davies, Hopkins,
  Baldry, Agius, Bauer, Bland-Hawthorn, Brough, Brown, Cluver, {De Propris},
  Drinkwater, Holwerda, Kelvin, Lara-Lopez, Liske, Lopez-Sanchez, Loveday,
  Mahajan, McNaught-Roberts, Moffett, Norberg, Obreschkow, Owers, Penny,
  Pimbblet, Prescott, Taylor, van Kampen, \& Wilkins}]{Robotham2014a}
Robotham A. S.~G. {et~al.}, 2014, MNRAS, 444, 3986

\bibitem[{Rodriguez-Gomez {et~al}\mbox{.}(2015)Rodriguez-Gomez, Genel,
  Vogelsberger, Sijacki, Pillepich, Sales, Torrey, Snyder, Nelson, Springel,
  Ma, \& Hernquist}]{Rodriguez-Gomez2015}
Rodriguez-Gomez V. {et~al.}, 2015, MNRAS, 449, 49

\bibitem[{Sales {et~al}\mbox{.}(2012)Sales, Navarro, Theuns, Schaye, White,
  Frenk, Crain, \& {Dalla Vecchia}}]{Sales2012}
Sales L.~V., Navarro J.~F., Theuns T., Schaye J., White S. D.~M., Frenk C.~S.,
  Crain R.~a., {Dalla Vecchia} C., 2012, MNRAS, 423, 1544

\bibitem[{Sales {et~al}\mbox{.}(2015)Sales, Vogelsberger, Genel, Torrey,
  Nelson, Rodriguez-Gomez, Wang, Pillepich, Sijacki, Springel, \&
  Hernquist}]{Sales2014a}
Sales L.~V. {et~al.}, 2015, MNRAS, 447, L6

\bibitem[{Searle \& Zinn(1978)}]{Searle1978}
Searle L., Zinn R., 1978, ApJ, 225, 357

\bibitem[{Sparre {et~al}\mbox{.}(2015)Sparre, Hayward, Springel, Vogelsberger,
  Genel, Torrey, Nelson, Sijacki, \& Hernquist}]{Sparre2015}
Sparre M. {et~al.}, 2015, MNRAS, 447, 3548

\bibitem[{Springel(2010)}]{Springel2010}
Springel V., 2010, MNRAS, 401, 791

\bibitem[{Springel {et~al}\mbox{.}(2001)Springel, White, Tormen, \&
  Kauffmann}]{Springel2001}
Springel V., White S. D.~M., Tormen G., Kauffmann G., 2001, MNRAS, 328, 726

\bibitem[{Stewart {et~al}\mbox{.}(2009)Stewart, Bullock, Barton, \&
  Wechsler}]{Stewart2009}
Stewart K.~R., Bullock J.~S., Barton E.~J., Wechsler R.~H., 2009, ApJ, 702,
  1005

\bibitem[{Tissera {et~al}\mbox{.}(2014)Tissera, Beers, Carollo, \&
  Scannapieco}]{Tissera2014}
Tissera P.~B., Beers T.~C., Carollo D., Scannapieco C., 2014, MNRAS, 439, 3128

\bibitem[{Tissera {et~al}\mbox{.}(2013)Tissera, Scannapieco, Beers, \&
  Carollo}]{Tissera2013}
Tissera P.~B., Scannapieco C., Beers T.~C., Carollo D., 2013, MNRAS, 432, 3391

\bibitem[{Tissera, White \& Scannapieco(2012)Tissera, White, \&
  Scannapieco}]{Tissera2012}
Tissera P.~B., White S. D.~M., Scannapieco C., 2012, MNRAS, 420, 255

\bibitem[{Torrey {et~al}\mbox{.}(2014)Torrey, Vogelsberger, Genel, Sijacki,
  Springel, \& Hernquist}]{Torrey2014}
Torrey P., Vogelsberger M., Genel S., Sijacki D., Springel V., Hernquist L.,
  2014, MNRAS, 438, 1985

\bibitem[{Torrey {et~al}\mbox{.}(2015)Torrey, Wellons, Machado, Griffen,
  Nelson, Rodriguez-Gomez, McKinnon, Pillepich, Ma, Vogelsberger, Springel, \&
  Hernquist}]{Torrey2015}
Torrey P. {et~al.}, 2015, MNRAS, 454, 2770

\bibitem[{Trujillo {et~al}\mbox{.}(2006)Trujillo, Feulner, Goranova, Hopp,
  Longhetti, Saracco, Bender, Braito, {Della Ceca}, Drory, Mannucci, \&
  Severgnini}]{Trujillo2006}
Trujillo I. {et~al.}, 2006, MNRAS, 373, L36

\bibitem[{van Dokkum {et~al}\mbox{.}(2008)van Dokkum, Franx, Kriek, Holden,
  Illingworth, Magee, Bouwens, Marchesini, Quadri, Rudnick, Taylor, \&
  Toft}]{VanDokkum2008}
van Dokkum P. {et~al.}, 2008, ApJ, 677, L5

\bibitem[{van Dokkum {et~al}\mbox{.}(2010)van Dokkum, Whitaker, Brammer, Franx,
  Kriek, Labb{\'{e}}, Marchesini, Quadri, Bezanson, Illingworth, Muzzin,
  Rudnick, Tal, \& Wake}]{VanDokkum2010}
van Dokkum P.~G. {et~al.}, 2010, ApJ, 709, 1018

\bibitem[{Vogelsberger {et~al}\mbox{.}(2013)Vogelsberger, Genel, Sijacki,
  Torrey, Springel, \& Hernquist}]{Vogelsberger2013}
Vogelsberger M., Genel S., Sijacki D., Torrey P., Springel V., Hernquist L.,
  2013, MNRAS, 436, 3031

\bibitem[{Vogelsberger {et~al}\mbox{.}(2014{\natexlab{a}})Vogelsberger, Genel,
  Springel, Torrey, Sijacki, Xu, Snyder, Bird, Nelson, \&
  Hernquist}]{Vogelsberger2014}
Vogelsberger M. {et~al.}, 2014{\natexlab{a}}, Nature, 509, 177

\bibitem[{Vogelsberger {et~al}\mbox{.}(2014{\natexlab{b}})Vogelsberger, Genel,
  Springel, Torrey, Sijacki, Xu, Snyder, Nelson, \&
  Hernquist}]{Vogelsberger2014a}
Vogelsberger M. {et~al.}, 2014{\natexlab{b}}, MNRAS, 444, 1518

\bibitem[{Vulcani {et~al}\mbox{.}(2015)Vulcani, Marchesini, {De Lucia}, Muzzin,
  Stefanon, Brammer, Labbe', \& Milvang-Jensen}]{Vulcani2015}
Vulcani B., Marchesini D., {De Lucia} G., Muzzin A., Stefanon M., Brammer
  G.~B., Labbe' I., Milvang-Jensen B., 2015, preprint (arXiv:1509.00486)

\bibitem[{Wellons {et~al}\mbox{.}(2016)Wellons, Torrey, Ma, Rodriguez-Gomez,
  Pillepich, Nelson, Genel, Vogelsberger, \& Hernquist}]{Wellons2016}
Wellons S. {et~al.}, 2016, MNRAS, 456, 1030

\bibitem[{Zavala {et~al}\mbox{.}(2012)Zavala, Avila-Reese, Firmani, \&
  Boylan-Kolchin}]{Zavala2012}
Zavala J., Avila-Reese V., Firmani C., Boylan-Kolchin M., 2012, MNRAS, 427,
  1503

\bibitem[{Zolotov {et~al}\mbox{.}(2009)Zolotov, Willman, Brooks, Governato,
  Brook, Hogg, Quinn, \& Stinson}]{Zolotov2009}
Zolotov A., Willman B., Brooks A.~M., Governato F., Brook C.~B., Hogg D.~W.,
  Quinn T., Stinson G., 2009, ApJ, 702, 1058

\bibitem[{Zolotov {et~al}\mbox{.}(2010)Zolotov, Willman, Brooks, Governato,
  Hogg, Shen, \& Wadsley}]{Zolotov2010}
Zolotov A., Willman B., Brooks A.~M., Governato F., Hogg D.~W., Shen S.,
  Wadsley J., 2010, ApJ, 721, 738

\end{thebibliography}

\end{document}